\documentstyle[epsf]{mn2e}

\newcommand{\sD}{{\cal D}}

\newcommand{\bF}{\mbox{\boldmath$F$}}

\title[Tidally distorted accretion discs]
{Tidally distorted accretion discs in binary stars}

\author[G. I. Ogilvie]
  {G. I. Ogilvie\\
  Institute of Astronomy, University of Cambridge, Madingley Road,
  Cambridge CB3 0HA}

\begin{document}

\maketitle

\label{firstpage}

\begin{abstract}
  The non-axisymmetric features observed in the discs of dwarf novae
  in outburst are usually considered to be spiral shocks, which are
  the non-linear relatives of tidally excited waves.  This
  interpretation suffers from a number of problems.  For example, the
  natural site of wave excitation lies outside the Roche lobe, the
  disc must be especially hot, and most treatments of wave propagation
  do not take into account the vertical structure of the disc.
  
  In this paper I construct a detailed semi-analytical model of the
  non-linear tidal distortion of a thin, three-dimensional accretion
  disc by a binary companion on a circular orbit.  The analysis
  presented here allows for vertical motion and radiative energy
  transport, and introduces a simple model for the turbulent magnetic
  stress.  The $m=2$ inner vertical resonance has an important
  influence on the amplitude and phase of the tidal distortion.  I
  show that the observed patterns find a natural explanation if the
  emission is associated with the tidally thickened sectors of the
  outer disc, which may be irradiated from the centre.  According to
  this hypothesis, it may be possible to constrain the physical
  parameters of the disc through future observations.
\end{abstract}

\begin{keywords}
  accretion, accretion discs -- binaries: close -- celestial mechanics
  -- hydrodynamics -- MHD -- turbulence.
\end{keywords}

\section{Introduction}

\subsection{Tidal effects on accretion discs}

Accretion discs are commonly found in interacting binary star systems
where matter is transferred from a normal star towards a compact
companion (e.g. Lubow \& Shu 1975).  Tidal forces have an important
role in these systems.  The binary orbit is usually made accurately
circular by the tidal interaction between the stars.  The disc
surrounding the compact star also experiences a strong tidal force
from the companion, resulting in a non-axisymmetric distortion of the
disc.  Angular momentum transported outwards through the disc by
`viscous' stresses is transferred to the binary orbit through tidal
torques.\footnote{In reality the `viscous' stress is most likely to be
  a turbulent magnetic stress resulting from the non-linear
  development of the magnetorotational instability (Balbus \& Hawley
  1998).}

The tidal interaction between a disc and an orbiting companion is also
important in young binary stars and protoplanetary systems (e.g. Lin
\& Papaloizou 1993).  In these situations there may be discs both
interior and exterior to the companion's orbit, which need not be
circular.

Early theoretical studies addressed the tidal influence of a companion
with a circular orbit and a mass ratio $q=M_2/M_1$ of order unity on a
circumstellar disc (Paczy\'nski 1977; Papaloizou \& Pringle 1977).
The method of Paczy\'nski (1977) makes use of the idea that, in a thin
accretion disc, the effects of pressure and viscosity are relatively
weak.  In the absence of resonances, and provided that the
trajectories do not intersect, the motion of the gas is essentially
ballistic.  The ballistic trajectories in a circular binary system are
just the orbits of the restricted three-body problem, which has been
studied in great detail in celestial mechanics (e.g. Sz\'ebehely
1967).  Paczy\'nski (1977) showed that there is a family of these
orbits that lie in the binary plane and are periodic in the binary
frame, enclosing the compact star (Fig.~1).  They reduce to circular
Keplerian orbits when the companion star is removed.  It is therefore
natural to suppose that the streamlines of a tidally distorted disc
correspond accurately to Paczy\'nski's orbits.  The maximum size of
the disc can be identified as the largest non-intersecting orbit.

\begin{figure*}
  \centerline{\epsfbox{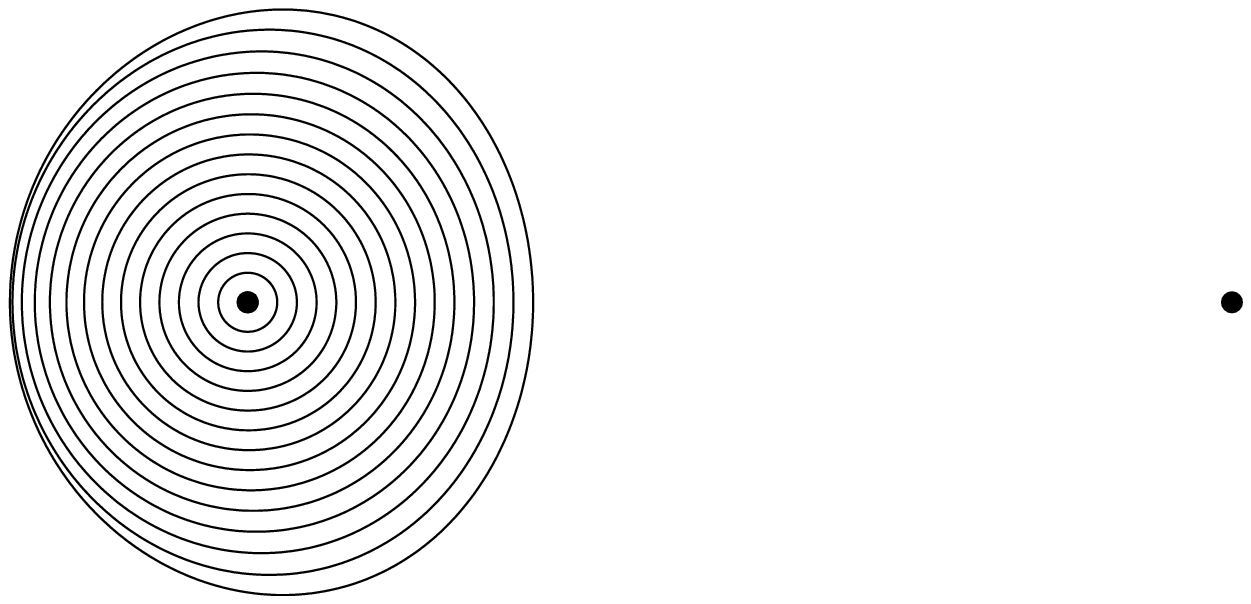}}
  \caption{Paczy\'nski's orbits in the restricted three-body
    problem for a mass ratio $q=1$.  The prograde circumstellar case
    is depicted.}
\end{figure*}

An alternative way of analysing the tidal distortion of a disc is
through linear perturbation theory.  A solution is first obtained for
an axisymmetric disc around an isolated star.  The tidal force is
introduced as a perturbation and the linearized equations are solved
to determine the tidal distortion.  A solution can be found that is
steady in the binary frame.  The distortion of the disc is very
slightly out of phase with the tidal forcing owing to the effective
viscosity of the disc.  This results in a tidal torque that increases
with increasing radius until it becomes comparable to the local
viscous torque and the disc is truncated.  Such a calculation was
carried out for a two-dimensional disc model by Papaloizou \& Pringle
(1977).  At least for mass ratios of order unity, the tidal truncation
radii determined by the two methods agree reasonably well.

\subsection{Resonances}

Additional complications arise if the disc contains radii at which the
tidal forcing resonates with a natural mode of oscillation of the
disc.  For mass ratios $q\la0.25$, a band of Paczy\'nski's orbits
around the 3:1 resonance are parametrically destabilized (Hirose \&
Osaki 1990).  In a continuous disc this causes a local growth of
eccentricity (Lubow 1991), which may be able to compete with
dissipative processes to sustain a global eccentric mode of the disc,
as observed in superhump binaries.

When the mass of the companion is much smaller, one or more Lindblad
resonances may be reached before the disc is truncated.  Such
resonances usually result in the launching of non-axisymmetric waves
(Goldreich \& Tremaine 1979).  Unless the waves are able to propagate
a significant distance through the disc, the torque exerted at a
Lindblad resonance is localized and independent of the viscosity.
Such a localized torque will often truncate the disc in the case of a
very low-mass companion (e.g. Lin \& Papaloizou 1986).

\subsection{Three-dimensional effects}

The methods of Paczy\'nski (1977) and Papaloizou \& Pringle (1977)
both neglect the three-dimensional nature of the disc.  In a linear
analysis, Lubow (1981) has shown that the vertical component of the
tidal force induces homogeneous vertical expansions and contractions
of the disc, which are characterized by a vertical velocity
proportional to the distance $z$ from the mid-plane.  The natural
frequency of this mode of oscillation, which corresponds to the $n=1$
p mode in the notation of Lubow \& Pringle (1993), or the ${\rm
  p}_1^{\rm e}$ mode in the classification of Ogilvie (1998), depends
on the adiabatic exponent $\gamma$.

As in the case of horizontal forcing, the response to vertical tidal
forcing may involve resonant and non-resonant components.  The
equivalent of a Lindblad resonance for the ${\rm p}_1^{\rm e}$ mode is
a vertical resonance\footnote{Distinct from the vertical resonances
  that excite bending modes (e.g. Shu, Cuzzi \& Lissauer 1983).  Here
  we are concerned with motions that are symmetric about the
  mid-plane.}, the location of which depends on $\gamma$ (Lubow 1981).
If such a resonance lies within the disc, a resonant torque is exerted
there and waves may be excited.  Vertical resonances in a thin disc in
a binary system are intrinsically much weaker than Lindblad
resonances.  However, for mass ratios of order unity, the $m=2$ inner
vertical resonance typically lies within the tidal truncation radius,
while the Lindblad resonances all lie well outside the disc (outside
the Roche lobe, in fact).  The vertical resonance can therefore have
an important influence on the disc.

There is also a non-resonant, non-wavelike response to vertical tidal
forcing at all radii.  Dissipation can result in a significant phase
shift in this response, causing an additional non-resonant tidal
torque.

Although these effects have been widely neglected in the literature,
Stehle \& Spruit (1999) have recently proposed an interesting method
of including vertical motion, in an approximate way, in
two-dimensional numerical simulations of accretion discs.  Stehle \&
Spruit (1999) also drew attention to the resonant excitation of
vertical oscillations by the $m=2$ component of the tidal potential.

\subsection{Non-axisymmetric structure in accretion discs}

In recent years, non-axisymmetric features have been detected in the
discs of several dwarf novae in outburst using Doppler tomography
(Steeghs, Harlaftis \& Horne 1997; Steeghs 2001 and references
therein).  The tomograms indicate the surface brightness of the disc,
in the selected emission line, in a two-dimensional velocity space.
The pattern generally appears as two crescents forming a broken ring,
and the same is true when the pattern is mapped into real space, if
circular Keplerian motion is assumed.

Steeghs et al. (1997) interpreted the non-axisymmetric features as
evidence of `spiral structure', `spiral waves' or `spiral shocks' in
the disc.  Observers of similar features in other systems have
confidently reported them as spiral shocks (Joergens, Spruit \& Rutten
2000; Groot 2001).  Nevertheless, owing to the substantial
complications in predicting line emission from cataclysmic variable
discs (e.g. Robinson, Marsh \& Smak 1993), these identifications
should be regarded with some caution.  The tomograms may not
accurately reflect the dynamics of the disc, because they originate in
the tenuous layers of the atmosphere.  It might even be questioned
whether the observed features indeed resemble either a spiral or a
wave.

Two-armed spiral shocks had previously been seen in two-dimensional
numerical simulations of the formation of very hot accretion discs by
Roche-lobe overflow (Sawada, Matsuda \& Hachisu 1986).  Spruit (1987)
showed with an analytical model that they could allow accretion to
occur in an essentially inviscid disc, although the accretion rate is
very small in a thin disc.  Spiral shocks in two-dimensional disc
models can be understood as the non-linear relatives of linear
acoustic--inertial waves (Larson 1990).  A popular conception is that
such waves are excited by tidal forcing at the outer edge of the disc
and steepen into shocks as they propagate inwards.  Although the $m=2$
inner Lindblad resonance lies outside the Roche lobe, it is sometimes
considered to launch waves through its `virtual' effect on the disc
(Savonije, Papaloizou \& Lin 1994).

However, an analysis of Lindblad resonances in three-dimensional discs
(Lubow \& Ogilvie 1998) indicates that the dominant wave that is
launched is the fundamental mode of even symmetry, or ${\rm f}^{\rm
  e}$ mode.  Only in a vertically isothermal disc does this mode
behave like the waves in a two-dimensional disc model; in a thermally
stratified disc the ${\rm f}^{\rm e}$ mode behaves like a surface
gravity mode and has quite different propagational properties.
Typically, it dissipates after travelling only a short distance from
the Lindblad resonance (Lubow \& Ogilvie 1998; Bate et al. 2001).

Moreover, as pointed out by Godon, Livio \& Lubow (1998), spiral waves
in a thin accretion disc tend to be tightly wound except near
resonances.  If the non-axisymmetric features are to be interpreted as
a two-armed wave, then the observations indicate at most one radial
wavelength.  Either the wave is highly localized in the radial
direction, or it is not tightly wound.  As the Lindblad resonances lie
well outside the Roche lobe in IP Peg, it is very difficult to account
for unwound waves in two-dimensional disc models unless the disc is
unrealistically hot.

\subsection{Vertical resonances}

However, as mentioned by Godon et al. (1998), a vertical resonance
could lie inside the disc in IP Peg.  In fact, the location of the
$m=2$ inner vertical resonance for an adiabatic exponent $\gamma=5/3$
agrees well with the observed features.  In an inviscid, linear theory
a ${\rm p}_1^{\rm e}$-mode wave is emitted from the resonance (Lubow
1981), which would not be tightly wound near the site of launching.
However, in a separate paper (Ogilvie, in preparation) I make a
detailed analysis of this resonance allowing for non-linearity and
dissipation, and show that the resonance is typically saturated,
broadened and damped to the extent that the emitted wave can be
neglected.  Therefore no such wave will be considered in the present
paper.  Nevertheless, the presence of the vertical resonance still has
an important role in determining the tidal distortion, even if a wave
is not generated.

\subsection{Plan of the paper}

The purpose of this paper is to determine the non-linear tidal
distortion of a thin, three-dimensional accretion disc by a binary
companion on a circular orbit.  The fluid dynamical equations are
solved in three dimensions by means of asymptotic expansions, using
only the fact that the disc is thin.  The resulting model is more
complete than any previous analysis.  Regarding the two-dimensional
aspects of the problem, the model of Papaloizou \& Pringle (1977) is
extended by making it non-linear, while the model of Paczy\'nski
(1977) is developed by including the effects of pressure and turbulent
stresses.  In addition, allowance is made for vertical motion,
radiative energy transport and stress relaxation.  That is, a
Maxwellian viscoelastic model of the turbulent stress is adopted,
which accounts for the non-zero relaxation time of the turbulence.

This approach is part of the programme of `continuum celestial
mechanics' that has already addressed the dynamics of warped discs
(Ogilvie 1999, 2000) and eccentric discs (Ogilvie 2001) using a
closely related method.  It provides a completely independent
alternative to direct numerical simulations.  The present method is
more naturally suited to thin, but fully vertically resolved, discs.
It also allows a more rapid solution, and is in some respects more
advanced in physical content, than present-day simulations.

Although the finer details of the solutions obtained are not at
present observationally testable, comparisons can be made at a simple
level between the non-axisymmetric structure in the computed models
and the features observed in Doppler tomograms.  It will be seen that
the observed patterns find a natural explanation if the emission is
associated with the tidally thickened sectors of the outer disc, which
may be irradiated from the centre.

The remainder of this paper is organized as follows. In Section~2
Paczy\'nski's orbits are used to define a orbital coordinate system
that is used in solving the fluid dynamical equations.  In Section~3
the model for the turbulent stress is explained.  The basic equations
are written in the orbital coordinate system in Section~4.  The
asymptotic development and solution of the equations is worked out in
Section~5, leading to a prediction for the non-axisymmetric distortion
of the disc.  Simple synthetic Doppler tomograms are computed and
compared with observations in Section~6.  A summary and discussion of
the results are found in Section~7.

\section{Orbital coordinates}

Consider a binary system with two stars in a circular orbit.  The
stars are approximated as spherical masses $M_1$ and $M_2$.  Let $a$
be the binary separation and
\begin{equation}
  \omega=\left[{{G(M_1+M_2)}\over{a^3}}\right]^{1/2}
\end{equation}
the binary orbital frequency.

Consider the binary frame rotating about the centre of mass with
angular velocity $\omega$, and let $(R,\phi,z)$ be cylindrical polar
coordinates such that star 1 (about which the disc orbits) is at the
origin, while star 2 has fixed coordinates $(a,0,0)$.  The axis of
rotation passes through the centre of mass and is parallel to the
$z$-axis.  The Roche potential
\begin{eqnarray}
  \lefteqn{\Phi=-GM_1(R^2+z^2)^{-1/2}}&\nonumber\\
  &&-GM_2(R^2+a^2-2aR\cos\phi+z^2)^{-1/2}\\
  &&-{{1}\over{2}}\omega^2R^2+GM_2a^{-2}R\cos\phi
\end{eqnarray}
includes the gravitational potentials of the two stars and the
centrifugal potential.

The analysis of a tidally distorted disc is simplified by introducing
non-orthogonal {\it orbital coordinates} $(\lambda,\phi,z)$ based on
Paczy\'nski's orbits, instead of cylindrical polar coordinates.  The
region of the binary plane filled by Paczy\'nski's orbits is covered
by a coordinate system $(\lambda,\phi)$, where $\lambda$ is a
quasi-radial coordinate that labels the orbits, and $\phi$ is the
azimuthal angle.  This naturally accounts for the principal tidal
distortion of the disc.  In this paper, only the case of a prograde
circumstellar disc is considered, although the generalization to
retrograde or circumbinary discs is straightforward.

There are many possible ways to label the orbits.  In the case of
eccentric discs (Ogilvie 2001) it was natural to use the semi-latus
rectum, which measures the angular momentum of the orbits, and reduces
to $R$ in the limit of circular orbits.  However, the angular momentum
is not constant on Paczy\'nski's orbits.  A simple alternative is to
define $\lambda$ to be the value of $R$ at $\phi=0$.  In the absence
of star 2, $\lambda$ reduces to $R$.  In general, there exists a
function $R(\lambda,\phi)$, satisfying $R(\lambda,0)=\lambda$, that
defines the shape of the orbits and must be determined numerically.

To carry out the vector calculus required in the fluid-dynamical
equations, one must obtain the metric coefficients and connection
components (cf. Ogilvie 2001).  The two-dimensional line element is
\begin{equation}
  {\rm d}s^2={\rm d}R^2+R^2\,{\rm d}\phi^2=
  (R_\lambda\,{\rm d}\lambda+R_\phi\,{\rm d}\phi)^2+R^2\,{\rm d}\phi^2,
\end{equation}
where the subscripts on $R$ stand for partial derivatives of the
function $R(\lambda,\phi)$.  Thus the metric coefficients are
\begin{equation}
  g_{\lambda\lambda}=R_\lambda^2,\qquad
  g_{\lambda\phi}=R_\lambda R_\phi,\qquad
  g_{\phi\phi}=R^2+R_\phi^2.
\end{equation}
As expected, the square root of the metric determinant is equal to the
Jacobian of the coordinate system:
\begin{equation}
  g^{1/2}=J={{\partial(x,y)}\over{\partial(\lambda,\phi)}}
  ={{\partial(x,y)}\over{\partial(R,\phi)}}
  {{\partial(R,\phi)}\over{\partial(\lambda,\phi)}}=RR_\lambda.
\end{equation}
The inverse metric coefficients are
\begin{equation}
  g^{\lambda\lambda}={{R^2+R_\phi^2}\over{R^2R_\lambda^2}},\qquad
  g^{\lambda\phi}=-{{R_\phi}\over{R^2R_\lambda}},\qquad
  g^{\phi\phi}={{1}\over{R^2}}.
\end{equation}
The components of the metric connection, given by
\begin{equation}
  \Gamma^a_{bc}={\textstyle{{1}\over{2}}}g^{ad}
  (\partial_b g_{cd}+\partial_c g_{db}-\partial_d g_{bc}),
\end{equation}
are found to be
\begin{equation}
  \Gamma^\lambda_{\lambda\lambda}=
  {{R_{\lambda\lambda}}\over{R_\lambda}},\qquad
  \Gamma^\lambda_{\lambda\phi}={{R_{\lambda\phi}}\over{R_\lambda}}-
  {{R_\phi}\over{R}},
\end{equation}
\begin{equation}
  \Gamma^\lambda_{\phi\phi}=-{{(R^2+2R_\phi^2-RR_{\phi\phi})}\over
  {RR_\lambda}},
\end{equation}
\begin{equation}
  \Gamma^\phi_{\lambda\lambda}=0,\qquad
  \Gamma^\phi_{\lambda\phi}={{R_\lambda}\over{R}},\qquad
  \Gamma^\phi_{\phi\phi}={{2R_\phi}\over{R}}.
\end{equation}

The coordinate system is trivially extended to three dimensions by
incorporating the vertical coordinate $z$.  With the exception of
$g_{zz}=g^{zz}=1$, all metric coefficients and connection components
involving $z$ vanish.  The three-dimensional Levi-Civita tensor
(alternating tensor) is
\begin{equation}
  \eta^{abc}=J^{-1}\epsilon_{abc},
\end{equation}
where $\epsilon_{abc}$ denotes the usual permutation symbol with
values $0,\pm1$.

\section{Modelling the turbulent stress}

It is now widely accepted that the turbulent stress in most, if not
all, accretion discs is magnetohydrodynamic (MHD) in origin, resulting
from the non-linear development of the magnetorotational instability
(Balbus \& Hawley 1998).  In MHD turbulence any fluid variable $X$
exhibits fluctuations $X'$ about a mean value $\langle X\rangle$, so
that $X=\langle X\rangle+X'$ with $\langle X'\rangle=0$.  Correlations
between fluctuating quantities result in a non-zero mean turbulent
Maxwell stress tensor,
\begin{equation}
  {{1}\over{4\pi}}\bigg\langle
  B_i'B_j'-{{1}\over{2}}B_k'B_k'\,\delta_{ij}\bigg\rangle,
\end{equation}
and Reynolds stress tensor,
\begin{equation}
  -\bigg\langle\rho u_i'u_j'+\rho'(u_iu_j'+u_i'u_j+u_i'u_j')\bigg\rangle.
\end{equation}
(In this section only, Cartesian tensor notation is used, together
with the summation convention.)  In numerical simulations the magnetic
contribution is found to dominate and therefore any model of the
stress should aim primarily to reflect the dynamics of the Maxwell
tensor.

In a recent paper (Ogilvie 2001) I proposed a simple viscoelastic
model of the turbulent stress in an accretion disc.  This was put
forward as a straightforward generalization of the standard viscous
model, used extensively in accretion disc theory, to incorporate the
effect of a non-zero relaxation time for the turbulence.  It was also
motivated by certain formal and physical analogies between MHD and
viscoelasticity.  Here I present a slightly refined version that
accounts more completely for the transfer of energy between the mean
motion and the turbulence, to heat and then to radiation.

The turbulent stress is written as
$T_{ij}-{\textstyle{{1}\over{2}}}T_{kk}\delta_{ij}$, where $T_{ij}$ is
a symmetric second-rank tensor that can be associated approximately
with $\langle B_i'B_j'\rangle/4\pi$.\footnote{The association is not
  perfect because equation (\ref{tij}) does not necessarily guarantee
  that $T_{ij}$ is a positive semi-definite tensor, as $\langle
  B_i'B_j'\rangle/4\pi$ should be.}  The equation of motion is then
\begin{equation}
  \rho(\partial_tu_i+u_j\partial_ju_i)=-\rho\partial_i\Phi-
  \partial_i(p+{\textstyle{{1}\over{2}}}T_{jj})+\partial_jT_{ij},
\end{equation}
where $\rho$ is the density, $u_i$ the velocity, $\Phi$ the external
gravitational potential and $p$ the pressure.  The tensor $T_{ij}$ is
taken to satisfy the equation
\begin{equation}
  T_{ij}+\tau\sD T_{ij}=\mu(\partial_iu_j+\partial_ju_i)-
  (2\mu_{\rm b}+{\textstyle{{2}\over{3}}}\mu)(\partial_ku_k)\delta_{ij},
  \label{tij}
\end{equation}
where $\tau$ is the relaxation time, $\mu$ the effective (dynamic)
viscosity and $\mu_{\rm b}$ the effective bulk viscosity.  The
operator $\sD$ is a convective derivative acting on second-rank tensors,
and is defined by
\begin{equation}
  \sD T_{ij}=(\partial_t+u_k\partial_k)T_{ij}-
  T_{ik}\partial_ku_j-T_{jk}\partial_ku_i+2T_{ij}\partial_ku_k.
\end{equation}

With the aid of the equation of mass conservation,
\begin{equation}
  \partial_t\rho+\partial_i(\rho u_i)=0,
\end{equation}
an equation for the kinetic and gravitational energy of the fluid is
then obtained in the form
\begin{eqnarray}
  \lefteqn{\partial_t\left[\rho({\textstyle{{1}\over{2}}}u^2+\Phi)\right]+
  \partial_i\left[\rho({\textstyle{{1}\over{2}}}u^2+\Phi)u_i\right]}
  &\nonumber\\
  &&=-u_i\partial_i(p+{\textstyle{{1}\over{2}}}T_{jj})+u_i\partial_jT_{ij},
\end{eqnarray}
provided that $\Phi$ is independent of $t$.  The turbulent energy
density is identified as ${\textstyle{{1}\over{2}}}T_{ii}$ (which can
be associated approximately with $\langle B_i'B_i'\rangle/8\pi$), and
this satisfies the equation
\begin{eqnarray}
  \lefteqn{\partial_t({\textstyle{{1}\over{2}}}T_{ii})+
  \partial_i(T_{jj}u_i-T_{ij}u_j)=
  u_i\partial_i({\textstyle{{1}\over{2}}}T_{jj})-u_i\partial_jT_{ij}}
  &\nonumber\\
  &&-{{1}\over{\tau}}({\textstyle{{1}\over{2}}}T_{ii}+
  3\mu_{\rm b}\partial_iu_i),
\end{eqnarray}
obtained by taking the trace of equation (\ref{tij}).  The first two
terms on the right-hand side appear with the opposite sign in the
preceding equation, and therefore represent the transfer of energy
between the mean motion and the turbulence.  The last term represents
the loss of turbulent energy to heat.  Accordingly the thermal energy
equation is written as
\begin{equation}
  \rho T(\partial_ts+u_i\partial_is)={{1}\over{\tau}}
  ({\textstyle{{1}\over{2}}}T_{ii}+3\mu_{\rm b}\partial_iu_i)-
  \partial_iF_i,
\end{equation}
where $T$ is the temperature, $s$ the specific entropy and $F_i$ the
radiative energy flux.  The total energy then satisfies an equation of
conservative form,
\begin{eqnarray}
  \lefteqn{\partial_t\left[\rho({\textstyle{{1}\over{2}}}u^2+\Phi+e)+
  {\textstyle{{1}\over{2}}}T_{ii}\right]}&\nonumber\\
  &&+\partial_i\left[\rho({\textstyle{{1}\over{2}}}u^2+\Phi+h)u_i+
  T_{jj}u_i-T_{ij}u_j+F_i\right]=0,
\end{eqnarray}
where $e$ is the specific internal energy, which satisfies the
thermodynamic relation ${\rm d}e=T\,{\rm d}s-p\,{\rm d}(1/\rho)$, and
$h=e+p/\rho$ is the specific enthalpy.

Consider the following two limits of the above system.  In the limit
$\tau\to0$ the stress is related instantaneously to the velocity
gradient by
\begin{equation}
  T_{ij}-{\textstyle{{1}\over{2}}}T_{kk}\delta_{ij}=
  \mu(\partial_iu_j+\partial_ju_i)+
  (\mu_{\rm b}-{\textstyle{{2}\over{3}}}\mu)(\partial_ku_k)\delta_{ij}.
\end{equation}
The equation of motion then reduces exactly to the compressible
Navier--Stokes equation.  In the limit $\tau\to\infty$, however, the
tensor $T_{ij}$ satisfies the equation
\begin{equation}
  \sD T_{ij}=0,
\end{equation}
which implies that $T_{ij}$ is `frozen in' to the fluid.  This is
precisely the equation satisfied by the tensor $B_iB_j/4\pi$ when the
magnetic field $B_i$ is `frozen in' to the fluid according to the
induction equation of ideal MHD.  The system of equations therefore
reduces exactly to ideal MHD in this limit.

For intermediate values of $\tau$, the viscoelastic model reproduces
both the viscous (dissipative) and elastic (magnetic) characteristics
of MHD turbulence.  It is intended to give a fair representation of
the dynamical response of the turbulent stress to a slowly varying
large-scale velocity field as found in a tidally distorted disc.  The
use of a tensor $T_{ij}$ in this model, rather than a vector or scalar
field, is essential in order to allow off-diagonal stress components.
Apart from the radiative energy flux, which in this paper is treated
in the diffusion approximation, the viscoelastic model for $\tau>0$
results in a hyperbolic system of equations and is therefore `causal'.
It also satisfies the continuum mechanical principle of material frame
indifference.

The viscoelastic model introduces a new dimensionless parameter for
the disc.  The Weissenberg number ${\rm We}$ can be defined as the
product of the relaxation time and the angular velocity, and is
expected to be of order unity, although it has never been measured
directly.  In the case of a circular, Keplerian disc the viscoelastic
model predicts the usual `viscous' stress component
$T_{R\phi}=-{\textstyle{{3}\over{2}}}\mu\Omega$ plus an additional
component $T_{\phi\phi}={\textstyle{{9}\over{2}}}{\rm We}\,\mu\Omega$.
If the effective viscosity is given by an alpha prescription,
$\mu=\alpha p/\Omega$, the quantity $T_{\phi\phi}$ can be associated
approximately with a toroidal magnetic field with magnetic pressure
${\textstyle{{9}\over{4}}}{\rm We}\,\alpha$ times the gas pressure.
The turbulent energy is stored in this component.  The shear energy is
tapped by the stress component $T_{R\phi}$, is converted into
turbulent energy stored in $T_{\phi\phi}$, and then goes into heat and
radiation.

\section{Basic equations}

\label{basic}

The basic equations governing a fluid disc in three dimensions are now
expressed in the orbital coordinate system.

The equation of mass conservation is
\begin{equation}
  (\partial_t+u^a\nabla_a)\rho=-\rho\nabla_au^a,
  \label{drho}
\end{equation}
where now $u^a$ is the velocity relative to the rotating frame.  The
equation of motion is
\begin{eqnarray}
  \lefteqn{\rho\left[(\partial_t+u^b\nabla_b)u^a+
  2\eta^{abc}\omega_bu_c\right]}&\nonumber\\
  &&=-\rho\nabla^a\Phi-\nabla^a(p+{\textstyle{{1}\over{2}}}{T^b}_b)+
  \nabla_bT^{ab},
  \label{du}
\end{eqnarray}
where $\omega^a$ is the angular velocity of the rotating frame (only
$\omega^z=\omega$ being non-zero).

As described above, the tensor $T^{ab}$ satisfies the equation
\begin{eqnarray}
  \lefteqn{T^{ab}+\tau\left[(\partial_t+u^c\nabla_c)T^{ab}-
  T^{ac}\nabla_cu^b-T^{bc}\nabla_cu^a+
  2T^{ab}\nabla_cu^c\right]}&\nonumber\\
  &&=\mu(\nabla^au^b+\nabla^bu^a)-
  (2\mu_{\rm b}+{\textstyle{{2}\over{3}}}\mu)(\nabla_cu^c)g^{ab},
  \label{tab}
\end{eqnarray}
which, like the induction equation in MHD, is not affected by the
rotation of the frame.  The energy equation for an ideal gas is
\begin{eqnarray}
  \lefteqn{\left({{1}\over{\gamma-1}}\right)(\partial_t+u^a\nabla_a)p=
  -\left({{\gamma}\over{\gamma-1}}\right)p\nabla_au^a}&\nonumber\\
  &&+{{1}\over{\tau}}({\textstyle{{1}\over{2}}}{T^a}_a+
  3\mu_{\rm b}\nabla_au^a)-\nabla_aF^a,
  \label{dp}
\end{eqnarray}
where $\gamma$ is the adiabatic exponent.  The radiative energy flux,
in the Rosseland approximation for an optically thick medium, is given
by
\begin{equation}
  F^a=-{{16\sigma T^3}\over{3\kappa\rho}}\nabla^aT,
\end{equation}
where $\sigma$ is the Stefan-Boltzmann constant and $\kappa$ the
opacity.  The equation of state of an ideal gas,
\begin{equation}
  p={{k\rho T}\over{\mu_{\rm m}m_{\rm H}}},
\end{equation}
is adopted, where $k$ is Boltzmann's constant, $\mu_{\rm m}$ the mean
molecular weight and $m_{\rm H}$ the mass of the hydrogen atom.  The
opacity is assumed to be of the Kramers form
\begin{equation}
  \kappa=C_\kappa\rho T^{-7/2},
\end{equation}
where $C_\kappa$ is a constant.  The generalization to other power-law
opacity functions is straightforward (Ogilvie 2000, 2001).

The effective viscosity coefficients are assumed to be given by an
alpha parametrization.  The precise form of the alpha prescription
relevant to a tidally distorted disc is to some extent debatable.  It
will be convenient to adopt the form
\begin{equation}
  \mu=\alpha p\left({{GM_1}\over{\lambda^3}}\right)^{-1/2},\qquad
  \mu_{\rm b}=\alpha_{\rm b}p\left({{GM_1}\over{\lambda^3}}\right)^{-1/2},
\end{equation}
where $\alpha$ and $\alpha_{\rm b}$ are the dimensionless shear and
bulk viscosity parameters.  In the limit of a circular disc in the
absence of star 2, this prescription reduces to the usual one,
$\mu=\alpha p/\tilde\Omega$, etc., where $\tilde\Omega$ is the
orbital angular velocity in the inertial frame.  Finally, the
Weissenberg number is defined by
\begin{equation}
  {\rm We}=\tau\left({{GM_1}\over{\lambda^3}}\right)^{1/2},
\end{equation}
where $\tau$ is the relaxation time.  Again, this reduces to ${\rm
  We}=\tilde\Omega\tau$ in the limit of a circular disc.  For
simplicity, the four dimensionless parameters of the disc,
$(\alpha,\alpha_{\rm b},{\rm We},\gamma)$, will be assumed to be
constant throughout this paper.  The other parameter in the problem is
the mass ratio $q=M_2/M_1$.

\section{Solution of the equations}

The analysis proceeds on the basis that the disc is almost steady in
the binary frame, and evolves only on a time-scale (the viscous
time-scale) that is long compared to the orbital time-scale.  It is
also assumed that the fluid variables vary smoothly in the horizontal
directions, not on a length-scale as short as the semi-thickness of
the disc.  Under these conditions, the temporal and spatial separation
of scales allows the fluid dynamical equations to be solved by
asymptotic methods.

Let the small parameter $\epsilon$ be a characteristic value of the
angular semi-thickness $H/R$ of the disc.  Adopt a system of units
such that the radius of the disc and the characteristic orbital
time-scale are $O(1)$.  Then define the stretched vertical coordinate
$\zeta=z/\epsilon$, which is $O(1)$ inside the disc.  The slow
evolution is captured by a slow time coordinate $T=\epsilon^2t$.

For the fluid variables, introduce the expansions
\begin{eqnarray}
  u^\lambda&=&\epsilon^2u^\lambda_2(\lambda,\phi,\zeta,T)+
  O(\epsilon^4),\\
  u^\phi&=&\Omega(\lambda,\phi)+
  \epsilon^2u^\phi_2(\lambda,\phi,\zeta,T)+O(\epsilon^4),\\
  u^z&=&\epsilon u^z_1(\lambda,\phi,\zeta,T)+O(\epsilon^3),\\
  \rho&=&\epsilon^s\left[\rho_0(\lambda,\phi,\zeta,T)+O(\epsilon^2)\right],\\
  p&=&\epsilon^{s+2}\left[p_0(\lambda,\phi,\zeta,T)+O(\epsilon^2)\right],\\
  T^{\lambda\lambda}&=&\epsilon^{s+2}\left[
  T^{\lambda\lambda}_0(\lambda,\phi,\zeta,T)+O(\epsilon^2)\right],\\
  T^{\lambda\phi}&=&\epsilon^{s+2}\left[
  T^{\lambda\phi}_0(\lambda,\phi,\zeta,T)+O(\epsilon^2)\right],\\
  T^{\phi\phi}&=&\epsilon^{s+2}\left[
  T^{\phi\phi}_0(\lambda,\phi,\zeta,T)+O(\epsilon^2)\right],\\
  T^{\lambda z}&=&\epsilon^{s+3}\left[
  T^{\lambda z}_1(\lambda,\phi,\zeta,T)+O(\epsilon^2)\right],\\
  T^{\phi z}&=&\epsilon^{s+3}\left[
  T^{\phi z}_1(\lambda,\phi,\zeta,T)+O(\epsilon^2)\right],\\
  T^{zz}&=&\epsilon^{s+2}\left[
  T^{zz}_0(\lambda,\phi,\zeta,T)+O(\epsilon^2)\right],\\
  \mu&=&\epsilon^{s+2}\left[\mu_0(\lambda,\phi,\zeta,T)+
  O(\epsilon^2)\right],\\
  \mu_{\rm b}&=&\epsilon^{s+2}\left[\mu_{{\rm b}0}(\lambda,\phi,\zeta,T)+
  O(\epsilon^2)\right],\\
  T&=&\epsilon^2\left[T_0(\lambda,\phi,\zeta,T)+O(\epsilon^2)\right],\\
  F^z&=&\epsilon^{s+3}\left[F_0(\lambda,\phi,\zeta,T)+
  O(\epsilon^2)\right],
\end{eqnarray}
while the horizontal components of $\bF$ are $O(\epsilon^{s+4})$.
Here $s$ is a positive parameter, which drops out of the analysis,
although formally one requires $s=11/3$ in order to balance
powers of $\epsilon$ in the opacity law.  Note that the dominant
motion is an orbital motion with angular velocity $\Omega$ in the
rotating frame, independent of $\zeta$ and $T$.

The Roche potential is expanded in a Taylor series about the
mid-plane,
\begin{equation}
  \Phi=\Phi_0(\lambda,\phi)+{\textstyle{{1}\over{2}}}\epsilon^2\zeta^2
  \Phi_2(\lambda,\phi)+O(\epsilon^4),
\end{equation}
where
\begin{eqnarray}
  \lefteqn{\Phi_0=-GM_1R^{-1}-
  GM_2(R^2+a^2-2aR\cos\phi)^{-1/2}}&\nonumber\\
  &&-{{1}\over{2}}\omega^2R^2+GM_2a^{-2}R\cos\phi,
\end{eqnarray}
\begin{equation}
  \Phi_2=GM_1R^{-3}+GM_2(R^2+a^2-2aR\cos\phi)^{-3/2}.
\end{equation}

When these expansions are substituted into the equations of
Section~\ref{basic}, various equations are obtained at different
orders in $\epsilon$.  The required equations comprise three sets,
representing three different physical problems, and these will be
considered in turn.

\subsection{Orbital motion}

\label{orbital}

The horizontal components of the equation of motion (\ref{du}) at
leading order [$O(\epsilon^s)$] are
\begin{equation}
  \rho_0\left(\Gamma^\lambda_{\phi\phi}\Omega^2-
  {{2g_{\phi\phi}}\over{J}}\omega\Omega\right)=
  -\rho_0(g^{\lambda\lambda}\partial_\lambda\Phi_0+
  g^{\lambda\phi}\partial_\phi\Phi_0),
\end{equation}
\begin{equation}
  \rho_0\left(\Omega\partial_\phi\Omega+
  \Gamma^\phi_{\phi\phi}\Omega^2+
  {{2g_{\lambda\phi}}\over{J}}\omega\Omega\right)=
  -\rho_0(g^{\lambda\phi}\partial_\lambda\Phi_0+
  g^{\phi\phi}\partial_\phi\Phi_0).
\end{equation}
As $\Phi_0$ is naturally given as a function of $R$ and $\phi$, these
may be written in the form
\begin{eqnarray}
  \lefteqn{(R^2+2R_\phi^2-RR_{\phi\phi})\Omega^2+
  2(R^2+R_\phi^2)\omega\Omega=}&\nonumber\\
  &&R\left({{\partial\Phi_0}\over{\partial R}}\right)_\phi-
  {{R_\phi}\over{R}}\left({{\partial\Phi_0}\over{\partial\phi}}\right)_R,
\end{eqnarray}
\begin{equation}
  \Omega\partial_\phi\left[R^2(\Omega+\omega)\right]=
  -\left({{\partial\Phi_0}\over{\partial\phi}}\right)_R.
  \label{am}
\end{equation}
Since no derivatives with respect to $\lambda$ appear, these equations
may be solved as a third-order system of ordinary differential
equations (ODEs) on $0\le\phi\le2\pi$ for each orbit separately.  It
is convenient to non-dimensionalize the equations by taking $a$ as the
unit of length and $(GM_1/a^3)^{1/2}$ as the unit of angular velocity.
The unknown functions $R$ and $\Omega$ are subject to periodic
boundary conditions.  The Jacobi constant,
\begin{equation}
  E={{1}\over{2}}(R^2+R_\phi^2)\Omega^2+\Phi_0,
  \label{jacobi}
\end{equation}
is an integral of the system, satisfying $E_\phi=0$.

The solution of these equations amounts to solving for Paczy\'nski's
orbits, and yields quantities such as $R$, $R_\phi$, $R_{\phi\phi}$,
$\Omega$ and $\partial_\phi\Omega$.  By differentiating the equations
with respect to $\lambda$, one obtains a further set of equations that
determine quantities such as $R_\lambda$, $R_{\lambda\phi}$ and
$\partial_\lambda\Omega$, which will all be required below.

\subsubsection{Surface density}

The equation of mass conservation (\ref{drho}) at leading order
[$O(\epsilon^s)$] is
\begin{equation}
  (\Omega\partial_\phi+u^z_1\partial_\zeta)\rho_0=
  -\rho_0\left[{{1}\over{J}}\partial_\phi(J\Omega)+\partial_\zeta u^z_1\right].
  \label{rho0a}
\end{equation}
Introduce the surface density at leading order [$O(\epsilon^{s+1})$],
\footnote{Throughout, integrals with respect to $\zeta$ are carried
  out over the full vertical extent of the disc, and integrals with
  respect to $\phi$ from $0$ to $2\pi$.}
\begin{equation}
  \tilde\Sigma(\lambda,\phi,T)=\int\rho_0\,{\rm d}\zeta.
\end{equation}
Then the vertically integrated version of equation (\ref{rho0a}) may
be written in the form
\begin{equation}
  \partial_\phi(J\tilde\Sigma\Omega)=0,
  \label{jsigom}
\end{equation}
which determines the variation of surface density around the orbit due
to the tidal distortion.  It will be convenient to introduce a
pseudo-circular surface density,
\begin{equation}
  \Sigma(\lambda,T)={{1}\over{2\pi\lambda}}
  \int J\tilde\Sigma\,{\rm d}\phi,
\end{equation}
which has the property that the mass contained between orbits
$\lambda_1$ and $\lambda_2$ is
\begin{equation}
  2\pi\int_{\lambda_1}^{\lambda_2}\Sigma\,\lambda\,{\rm d}\lambda.
\end{equation}

\subsubsection{Kinematic quantities}

The shear tensor of the orbital motion will be required below.  This
is defined in general by
\begin{equation}
  S^{ab}={\textstyle{{1}\over{2}}}(\nabla^au^b+\nabla^bu^a),
\end{equation}
and has the expansion
\begin{equation}
  S^{ab}=S^{ab}_0+O(\epsilon).
\end{equation}
The horizontal components at leading order are
\begin{equation}
  S^{\lambda\lambda}_0={{(R^3R_{\lambda\phi}+RR_{\lambda\phi}R_\phi^2+
  R_\lambda R_\phi^3-RR_\lambda R_\phi R_{\phi\phi})\Omega}\over
  {R^3R_\lambda^3}},
\end{equation}
\begin{eqnarray}
  \lefteqn{S^{\lambda\phi}_0={{(R^2+R_\phi^2)}\over{2R^2R_\lambda^2}}
  \partial_\lambda\Omega-
  {{R_\phi}\over{2R^2R_\lambda}}\partial_\phi\Omega}&\nonumber\\
  &&+{{(RR_\lambda R_{\phi\phi}-RR_\phi R_{\lambda\phi}-
  2R_\lambda R_\phi^2)\Omega}\over{2R^3R_\lambda^2}},
\end{eqnarray}
\begin{equation}
  S^{\phi\phi}_0=-{{R_\phi}\over{R^2R_\lambda}}\partial_\lambda\Omega+
  {{1}\over{R^2}}\partial_\phi\Omega+{{R_\phi\Omega}\over{R^3}}.
\end{equation}
The orbital contribution to the divergence $\nabla_au^a$ is
\begin{equation}
  \Delta={{1}\over{J}}\partial_\phi(J\Omega)=
  \left({{GM_1}\over{\lambda^3}}\right)^{1/2}g_1,
\end{equation}
where $g_1(\phi;\lambda)$ is a dimensionless quantity, equal to zero
for a circular disc, and which can be evaluated numerically from the
solution of the problem in Section~\ref{orbital}.

Note, however, that there is also a vertical shear component
$S^{zz}_0=\partial_\zeta u^z_1$ of the same order, which remains to be
evaluated.  This contributes to both the divergence and the
dissipation rate at leading order.

Let
\begin{equation}
  \Omega=\left({{GM_1}\over{\lambda^3}}\right)^{1/2}g_\Omega,
\end{equation}
\begin{equation}
  \omega=\left({{GM_1}\over{\lambda^3}}\right)^{1/2}g_\omega,
\end{equation}
where $g_\Omega(\phi;\lambda)$ and $g_\omega(\lambda)$ are
dimensionless quantities.  In the limit of a circular disc,
$g_\Omega+g_\omega=1$.

Noting that
\begin{equation}
  P=\int{{{\rm d}\phi}\over{\Omega}}
\end{equation}
is the orbital period, one may write
\begin{equation}
  \tilde\Sigma=\Sigma\,g_2,
\end{equation}
where $g_2(\phi;\lambda)$ is a dimensionless quantity, equal to unity
for a circular disc, and satisfying
\begin{equation}
  g_2={{2\pi\lambda}\over{J\Omega P}},
\end{equation}
\begin{equation}
  g_\Omega\partial_\phi\ln g_2=-g_1.
\end{equation}
Further useful quantities are defined according to
\begin{equation}
  \lambda\partial_\lambda\Omega=
  \left({{GM_1}\over{\lambda^3}}\right)^{1/2}\,g_3,
\end{equation}
\begin{equation}
  \partial_\phi\Omega=\left({{GM_1}\over{\lambda^3}}\right)^{1/2}\,g_4,
\end{equation}
\begin{equation}
  \Phi_2=\left({{GM_1}\over{\lambda^3}}\right)\,g_5
\end{equation}
and
\begin{equation}
  S^{ab}_0=\left({{GM_1}\over{\lambda^3}}\right)^{1/2}\,s^{ab},
\end{equation}
so that $(g_3,g_4,g_5,s^{\lambda\lambda},\lambda
s^{\lambda\phi},\lambda^2s^{\phi\phi})$ are dimensionless.

\subsection{Vertical structure and vertical motion}

\label{vertical}

The equation of mass conservation (\ref{drho}) at leading order
[$O(\epsilon^s)$] is
\begin{equation}
  (\Omega\partial_\phi+u^z_1\partial_\zeta)\rho_0=
  -\rho_0(\Delta+\partial_\zeta u^z_1).
  \label{rho0b}
\end{equation}
The energy equation (\ref{dp}) at leading order [$O(\epsilon^{s+2})$] is
\begin{eqnarray}
  \lefteqn{\left({{1}\over{\gamma-1}}\right)
  (\Omega\partial_\phi+u^z_1\partial_\zeta)p_0=
  -\left({{\gamma}\over{\gamma-1}}\right)p_0
  (\Delta+\partial_\zeta u^z_1)}&\nonumber\\
  &&+{{1}\over{2\tau}}\left[g_{\lambda\lambda}T^{\lambda\lambda}_0+
  2g_{\lambda\phi}T^{\lambda\phi}_0+g_{\phi\phi}T^{\phi\phi}_0+T^{zz}_0
  \right.\nonumber\\
  &&\left.\qquad\quad+6\mu_{{\rm b}0}(\Delta+\partial_\zeta u^z_1)\right]-
  \partial_\zeta F_0.
  \label{p0}
\end{eqnarray}
The vertical component of the equation of motion (\ref{du}) at
leading order [$O(\epsilon^{s+1})$] is
\begin{eqnarray}
  \lefteqn{\rho_0(\Omega\partial_\phi+u^z_1\partial_\zeta)u^z_1=
  -\rho_0\Phi_2\zeta}&\nonumber\\
  &&-\partial_\zeta\left[p_0+{\textstyle{{1}\over{2}}}
  \left(g_{\lambda\lambda}T^{\lambda\lambda}_0+
  2g_{\lambda\phi}T^{\lambda\phi}_0+g_{\phi\phi}T^{\phi\phi}_0-
  T^{zz}_0\right)\right].
  \label{w1}
\end{eqnarray}
The required components of the stress equation (\ref{tab}) at leading
order [$O(\epsilon^{s+2})$] are
\begin{eqnarray}
  \lefteqn{T^{\lambda\lambda}_0+\tau\left[
  (\Omega\partial_\phi+u^z_1\partial_\zeta)T^{\lambda\lambda}_0+
  2(\Delta+\partial_\zeta u^z_1)T^{\lambda\lambda}_0\right]}&\nonumber\\
  &&=2\mu_0S^{\lambda\lambda}_0-
  (2\mu_{{\rm b}0}+{\textstyle{{2}\over{3}}}\mu_0)
  (\Delta+\partial_\zeta u^z_1)g^{\lambda\lambda},
  \label{tll}
\end{eqnarray}
\begin{eqnarray}
  \lefteqn{T^{\lambda\phi}_0+\tau\left[
  (\Omega\partial_\phi+u^z_1\partial_\zeta)T^{\lambda\phi}_0-
  T^{\lambda\lambda}_0\partial_\lambda\Omega-
  T^{\lambda\phi}_0\partial_\phi\Omega\right.}&\nonumber\\
  &&\left.\qquad+2(\Delta+\partial_\zeta u^z_1)T^{\lambda\phi}_0\right]
  \nonumber\\
  &&=2\mu_0S^{\lambda\phi}_0-
  (2\mu_{{\rm b}0}+{\textstyle{{2}\over{3}}}\mu_0)
  (\Delta+\partial_\zeta u^z_1)g^{\lambda\phi},
\end{eqnarray}
\begin{eqnarray}
  \lefteqn{T^{\phi\phi}_0+\tau\left[
  (\Omega\partial_\phi+u^z_1\partial_\zeta)T^{\phi\phi}_0-
  2T^{\lambda\phi}_0\partial_\lambda\Omega-
  2T^{\phi\phi}_0\partial_\phi\Omega\right.}&\nonumber\\
  &&\left.\qquad+2(\Delta+\partial_\zeta u^z_1)T^{\phi\phi}_0\right]\nonumber\\
  &&=2\mu_0S^{\phi\phi}_0-
  (2\mu_{{\rm b}0}+{\textstyle{{2}\over{3}}}\mu_0)
  (\Delta+\partial_\zeta u^z_1)g^{\phi\phi},
\end{eqnarray}
\begin{eqnarray}
  \lefteqn{T^{zz}_0+\tau\left[
  (\Omega\partial_\phi+u^z_1\partial_\zeta)T^{zz}_0+
  2\Delta T^{zz}_0\right]}&\nonumber\\
  &&=2\mu_0\partial_\zeta u^z_1-
  (2\mu_{{\rm b}0}+{\textstyle{{2}\over{3}}}\mu_0)
  (\Delta+\partial_\zeta u^z_1).
  \label{tzz}
\end{eqnarray}
The constitutive relations at leading
order are
\begin{equation}
  F_0=-{{16\sigma T_0^{13/2}}\over{3C_\kappa\rho_0^2}}\partial_\zeta T_0,
  \label{f0}
\end{equation}
\begin{equation}
  p_0={{k\rho_0T_0}\over{\mu_{\rm m}m_{\rm H}}},
\end{equation}
\begin{equation}
  \mu_0=\alpha p_0\left({{GM_1}\over{\lambda^3}}\right)^{-1/2},\qquad
  \mu_{{\rm b}0}=\alpha_{\rm b}p_0\left({{GM_1}\over{\lambda^3}}\right)^{-1/2}.
  \label{mu0}
\end{equation}

The solution of equations (\ref{rho0b})--(\ref{mu0}) is found by a
non-linear separation of variables, a similar method to that used for
warped discs (Ogilvie 2000) and eccentric discs (Ogilvie 2001).  As an
intermediate step, one proposes that the solution should satisfy the
generalized vertical equilibrium relations
\begin{eqnarray}
  {{\partial p_0}\over{\partial\zeta}}&=&-f_2\,
  \rho_0\left({{GM_1}\over{\lambda^3}}\right)\zeta,
  \label{dpdz}\\
  {{\partial F_0}\over{\partial\zeta}}&=&f_1\,
  {{9\alpha}\over{4}}p_0\left({{GM_1}\over{\lambda^3}}\right)^{1/2},
\end{eqnarray}
in addition to equations (\ref{f0})--(\ref{mu0}).  Here
$f_1(\phi;\lambda)$ and $f_2(\phi;\lambda)$ are dimensionless
functions to be determined, and which are equal to unity for a
circular disc with ${\rm We}\to0$.  Physically, $f_1$ differs from
unity in a tidally distorted disc because of the enhanced dissipation
of energy and because of compressive heating and cooling (equation
\ref{p0}).  Similarly, $f_2$ reflects changes in the usual hydrostatic
vertical equilibrium resulting from the vertical velocity, the
turbulent stress and the variation of the vertical oscillation
frequency around the orbit (equation \ref{w1}).  Following Ogilvie
(2000), one identifies a natural physical unit for the thickness of
the disc,
\begin{eqnarray}
  \lefteqn{U_H=\left({{9\alpha}\over{4}}\right)^{1/14}\Sigma^{3/14}
  \left({{GM_1}\over{\lambda^3}}\right)^{-3/7}
  \left({{\mu_{\rm m}m_{\rm H}}\over{k}}\right)^{-15/28}}&\nonumber\\
  &&\times\left({{16\sigma}\over{3C_\kappa}}\right)^{-1/14},
\end{eqnarray}
and for the other variables according to
\begin{equation}
  U_\rho=\Sigma U_H^{-1},
\end{equation}
\begin{equation}
  U_p=\left({{GM_1}\over{\lambda^3}}\right)U_H^2U_\rho,
\end{equation}
\begin{equation}
  U_T=\left({{GM_1}\over{\lambda^3}}\right)
  \left({{\mu_{\rm m}m_{\rm H}}\over{k}}\right)U_H^2,
\end{equation}
\begin{equation}
  U_F=\left({{9\alpha}\over{4}}\right)
  \left({{GM_1}\over{\lambda^3}}\right)^{1/2}U_HU_p.
\end{equation}
The solution of the generalized vertical equilibrium equations is then
\begin{equation}
  \zeta=\zeta_*\,f_1^{1/14}f_2^{-13/28}g_2^{3/14}U_H,
  \label{solution_zeta}
\end{equation}
\begin{equation}
  \rho_0=\rho_*(\zeta_*)\,f_1^{-1/14}f_2^{13/28}g_2^{11/14}U_\rho,
  \label{solution_rho0}
\end{equation}
\begin{equation}
  p_0=p_*(\zeta_*)\,f_1^{1/14}f_2^{15/28}g_2^{17/14}U_p,
\end{equation}
\begin{equation}
  T_0=T_*(\zeta_*)\,f_1^{1/7}f_2^{1/14}g_2^{3/7}U_T,
\end{equation}
\begin{equation}
  F_0=F_*(\zeta_*)\,f_1^{8/7}f_2^{1/14}g_2^{10/7}U_F,
  \label{solution_f0}
\end{equation}
where the starred variables satisfy the dimensionless ODEs
\begin{eqnarray}
  {{{\rm d}p_*}\over{{\rm d}\zeta_*}}&=&-\rho_*\zeta_*,\\
  {{{\rm d}F_*}\over{{\rm d}\zeta_*}}&=&p_*,\\
  {{{\rm d}T_*}\over{{\rm d}\zeta_*}}&=&-\rho_*^2T_*^{-13/2}F_*,\\
  p_*&=&\rho_*T_*,
\end{eqnarray}
subject to the boundary conditions
\begin{equation}
  F_*(0)=\rho_*(\zeta_{{\rm s}*})=T_*(\zeta_{{\rm s}*})=0
\end{equation}
and the normalization of surface density,
\begin{equation}
  \int_{-\zeta_{{\rm s}*}}^{\zeta_{{\rm s}*}}\rho_*\,{\rm d}\zeta_*=1.
\end{equation}
Here $\zeta_{{\rm s}*}$ is the dimensionless height of the upper
surface of the disc.  The solution of these dimensionless ODEs is
easily obtained numerically in a once-for-all calculation (Ogilvie
2000), and one finds $\zeta_{{\rm s}*}\approx2.543$.  However, the
details of the solution do not affect the present
analysis.\footnote{It is assumed here that the disc is highly
  optically thick so that `zero boundary conditions' are adequate.
  The corrections for finite optical depth are described by Ogilvie
  (2000).}

One further proposes that the vertical velocity is of the form
\begin{equation}
  u^z_1=f_3\,\left({{GM_1}\over{\lambda^3}}\right)^{1/2}\zeta,
\end{equation}
where $f_3(\phi;\lambda)$ is a third dimensionless function, equal
to zero for a circular disc, and that the stress components at leading
order are of the form
\begin{equation}
  T^{ab}_0=t^{ab}(\phi;\lambda)\,p_0,
\end{equation}
so that $(t^{\lambda\lambda},\lambda
t^{\lambda\phi},\lambda^2t^{\phi\phi},t^{zz})$ are dimensionless
coefficients.

When these tentative solutions are substituted into equations
(\ref{rho0b})--(\ref{tzz}) one obtains a number of dimensionless ODEs in
$\phi$, which must be satisfied if the solution is to be valid.  From
equation (\ref{rho0b}) one obtains
\begin{equation}
  g_\Omega\left(-\partial_\phi\ln f_1+
  {{13}\over{2}}\partial_\phi\ln f_2\right)=-14f_3-3g_1.
  \label{ode1}
\end{equation}
Similarly, from equation (\ref{p0}) one obtains
\begin{eqnarray}
  \lefteqn{g_\Omega\left({{1}\over{\gamma-1}}\right)
  \partial_\phi\ln f_2=-\left({{\gamma+1}\over{\gamma-1}}\right)f_3
  -g_1}&
  \nonumber\\
  &&+{{1}\over{2{\rm We}}}\left[g_{\lambda\lambda}t^{\lambda\lambda}+
  2g_{\lambda\phi}t^{\lambda\phi}+g_{\phi\phi}t^{\phi\phi}+t^{zz}+
  6\alpha_{\rm b}(g_1+f_3)\right]\nonumber\\
  &&-{{9}\over{4}}\alpha f_1.
\end{eqnarray}
From equation (\ref{w1}) one obtains
\begin{eqnarray}
  \lefteqn{g_\Omega\partial_\phi f_3=-f_3^2-g_5}&\nonumber\\
  &&+f_2\left[1+{\textstyle{{1}\over{2}}}
  (g_{\lambda\lambda}t^{\lambda\lambda}+2g_{\lambda\phi}t^{\lambda\phi}+
  g_{\phi\phi}t^{\phi\phi}-t^{zz})\right].
\end{eqnarray}
Finally, from equations (\ref{tll})--(\ref{tzz}) one obtains
\begin{eqnarray}
  \lefteqn{t^{\lambda\lambda}+{\rm We}\left[
  g_\Omega(\partial_\phi t^{\lambda\lambda}+
  t^{\lambda\lambda}\partial_\phi\ln f_2)+
  (3f_3+g_1)t^{\lambda\lambda}\right]}&\nonumber\\
  &&=2\alpha s^{\lambda\lambda}-
  (2\alpha_{\rm b}+{\textstyle{{2}\over{3}}}\alpha)(g_1+f_3)g^{\lambda\lambda},
\end{eqnarray}
\begin{eqnarray}
  \lefteqn{\lambda t^{\lambda\phi}+{\rm We}\left[
  g_\Omega(\partial_\phi(\lambda t^{\lambda\phi})+
  \lambda t^{\lambda\phi}\partial_\phi\ln f_2)\right.}&\nonumber\\
  &&\left.\qquad\qquad+(3f_3+g_1)\lambda t^{\lambda\phi}-g_3t^{\lambda\lambda}-
  g_4\lambda t^{\lambda\phi}\right]\nonumber\\
  &&=2\alpha\lambda s^{\lambda\phi}-
  (2\alpha_{\rm b}+{\textstyle{{2}\over{3}}}\alpha)(g_1+f_3)
  \lambda g^{\lambda\phi},
\end{eqnarray}
\begin{eqnarray}
  \lefteqn{\lambda^2t^{\phi\phi}+{\rm We}\left[
  g_\Omega(\partial_\phi(\lambda^2t^{\phi\phi})+
  \lambda^2t^{\phi\phi}\partial_\phi\ln f_2)\right.}&\nonumber\\
  &&\left.\qquad\qquad+(3f_3+g_1)\lambda^2t^{\phi\phi}-
  2g_3\lambda t^{\lambda\phi}-
  2g_4\lambda^2t^{\phi\phi}\right]\nonumber\\
  &&=2\alpha\lambda^2s^{\phi\phi}-
  (2\alpha_{\rm b}+{\textstyle{{2}\over{3}}}\alpha)(g_1+f_3)
  \lambda^2g^{\phi\phi},
\end{eqnarray}
\begin{eqnarray}
  \lefteqn{t^{zz}+{\rm We}\left[
  g_\Omega(\partial_\phi t^{zz}+
  t^{zz}\partial_\phi\ln f_2)+(f_3+g_1)t^{zz}\right]}&\nonumber\\
  &&=2\alpha f_3-
  (2\alpha_{\rm b}+{\textstyle{{2}\over{3}}}\alpha)(g_1+f_3).
  \label{ode7}
\end{eqnarray}

These ODEs should be solved numerically for the functions
$(f_1,f_2,f_3,t^{\lambda\lambda},\lambda
t^{\lambda\phi},\lambda^2t^{\phi\phi},t^{zz})$ subject to periodic
boundary conditions $f_1(2\pi;\lambda)=f_1(0;\lambda)$, etc.  Note
that, in the limit of a circular disc ($g_1=0$, $g_2=1$,
$g_3=-{\textstyle{{3}\over{2}}}$, $g_4=0$, $g_5=1$,
$g^{\lambda\lambda}=1$, $\lambda g^{\lambda\phi}=0$,
$\lambda^2g^{\phi\phi}=1$, $s_{\lambda\lambda}=0$,
$\lambda^{-1}s_{\lambda\phi}=-{\textstyle{{3}\over{4}}}$,
$\lambda^{-2}s_{\phi\phi}=0$), the solution is $f_1=1$,
$f_2=(1+{\textstyle{{9}\over{4}}}{\rm We}\,\alpha)^{-1}$, $f_3=0$,
$t^{\lambda\lambda}=0$, $\lambda
t^{\lambda\phi}=-{\textstyle{{3}\over{2}}}\alpha$,
$\lambda^2t^{\phi\phi}={\textstyle{{9}\over{2}}}{\rm We}\,\alpha$,
$t^{zz}=0$.

\subsection{Slow velocities and time-evolution}

\label{slow}

It is possible to expand further in order to determine the mean
accretion flow in the disc and deduce the equation governing the
evolution of the pseudo-circular surface density $\Sigma(\lambda,T)$.
This is a generalization of the diffusion equation for the surface
density of a standard, circular accretion disc, which includes the
effects of tidal distortion and tidal torques.  The tidal truncation
of the disc should be discussed in the context of this evolutionary
equation.

The method is again similar to that used for eccentric discs (Ogilvie
2001), but is complicated by the fact that the specific angular
momentum is not constant on Paczy\'nski's orbits.  As this aspect of
the problem introduces much additional complexity without affecting
the main conclusions of this paper, it is deferred for future work.

\section{Investigation of the solutions}

\subsection{General remarks}

The preceding analysis is implemented numerically as follows.  First,
one selects the binary mass ratio $q$ and the dimensionless disc
parameters $(\alpha,\alpha_{\rm b},{\rm
  We},\gamma)$.\footnote{There is no difficulty in principle in
  extending the analysis of this paper to situations in which these
  dimensionless parameters vary within the disc.} A discrete set of
Paczy\'nski's orbits is computed, with equally spaced values of
$\lambda$ from the inner radius of the disc to the first point of
orbital intersection.  The dimensionless ODEs
(\ref{ode1})--(\ref{ode7}) are then solved to yield the functions
$f_i$ for each orbit.

The solutions may be interpreted by considering physical quantities of
particular interest, such as the surface brightness of the disc and
its semi-thickness.  The surface brightness (in continuum radiation)
is the value of the vertical radiative flux $F^z$ at the upper
surface.  From equation (\ref{solution_f0}) this may be written in the form
\begin{equation}
  F^+=f_F\,C_F\alpha^{8/7}
  \left({{GM_1}\over{\lambda^3}}\right)^{9/14}\Sigma^{10/7},
  \label{f+}
\end{equation}
where
\begin{equation}
  C_F=\left({{9}\over{4}}\right)^{8/7}
  F_*(\zeta_{{\rm s}*})
  \left({{\mu_{\rm m}m_{\rm H}}\over{k}}\right)^{-15/14}
  \left({{16\sigma}\over{3C_\kappa}}\right)^{-1/7}
\end{equation}
is a constant depending only on the opacity law, and
$f_F(\phi;\lambda)$ is a dimensionless function defined by
\begin{equation}
  f_F=f_1^{8/7}f_2^{1/14}g_2^{10/7},
\end{equation}
and equal to unity for a circular disc with ${\rm We}\to0$.  This
contains the azimuthal variation of the surface brightness and
represents the principal tidal distortion of the disc.  It does not
depend on $\Sigma(\lambda,T)$.  The additional variation of $F^+$ with
$\lambda$ and $T$ is contained in the remaining factors in equation
(\ref{f+}), and requires a knowledge of $\Sigma(\lambda,T)$.

Similarly, the semi-thickness of the disc is
\begin{equation}
  H=f_H\,C_H\alpha^{1/14}
  \left({{GM_1}\over{\lambda^3}}\right)^{-3/7}\Sigma^{3/14},
  \label{h}
\end{equation}
where
\begin{equation}
  C_H=\left({{9}\over{4}}\right)^{1/14}
  \zeta_{{\rm s}*}
  \left({{\mu_{\rm m}m_{\rm H}}\over{k}}\right)^{-15/28}
  \left({{16\sigma}\over{3C_\kappa}}\right)^{-1/14}
\end{equation}
is a constant depending only on the opacity law, and
$f_H(\phi;\lambda)$ is a dimensionless function defined by
\begin{equation}
  f_H=f_1^{1/14}f_2^{-13/28}g_2^{3/14},
  \label{fh}
\end{equation}
and equal to unity for a circular disc with ${\rm We}\to0$.

There are therefore two aspects to a complete solution for a tidally
distorted disc.  The first aspect concerns the determination of
Paczy\'nski's orbits and the functions $f_i$, which indicate how the
physical quantities vary around each orbit.  This part of the problem
does not depend on the distribution $\Sigma(\lambda,T)$ of surface
density over the set of orbits.  In this sense, the non-axisymmetric
tidal distortion of the disc is fixed and depends only on $q$ and the
dimensionless disc parameters $(\alpha,\alpha_{\rm b},{\rm
  We},\gamma)$.  However, to obtain the correct variation of physical
quantities from one orbit to the next, and in time, requires a
knowledge of $\Sigma(\lambda,T)$.  This second aspect of the problem
requires a solution of the evolutionary equation described briefly in
Section~\ref{slow}.  During the course of a dwarf-nova outburst, the
first aspect of the problem remains fixed except inasmuch as $\alpha$
varies, but the second aspect changes significantly as the surface
density evolves.

\subsection{Non-axisymmetric structure}

The non-axisymmetric structure of a tidally distorted disc may be
illustrated by plotting the functions $f_F(\phi;\lambda)$ and
$f_H(\phi;\lambda)$.  This is done in Fig.~2 for a binary mass ratio
$q=0.5$  and illustrative disc parameters $\alpha=0.1$, $\alpha_{\rm
  b}=0$, ${\rm We}=0.5$, and $\gamma=5/3$.  This represents a disc
with a Kramers opacity law and a moderate viscosity and relaxation
time.

\begin{figure*}
  \centerline{\epsfysize8cm\epsfbox{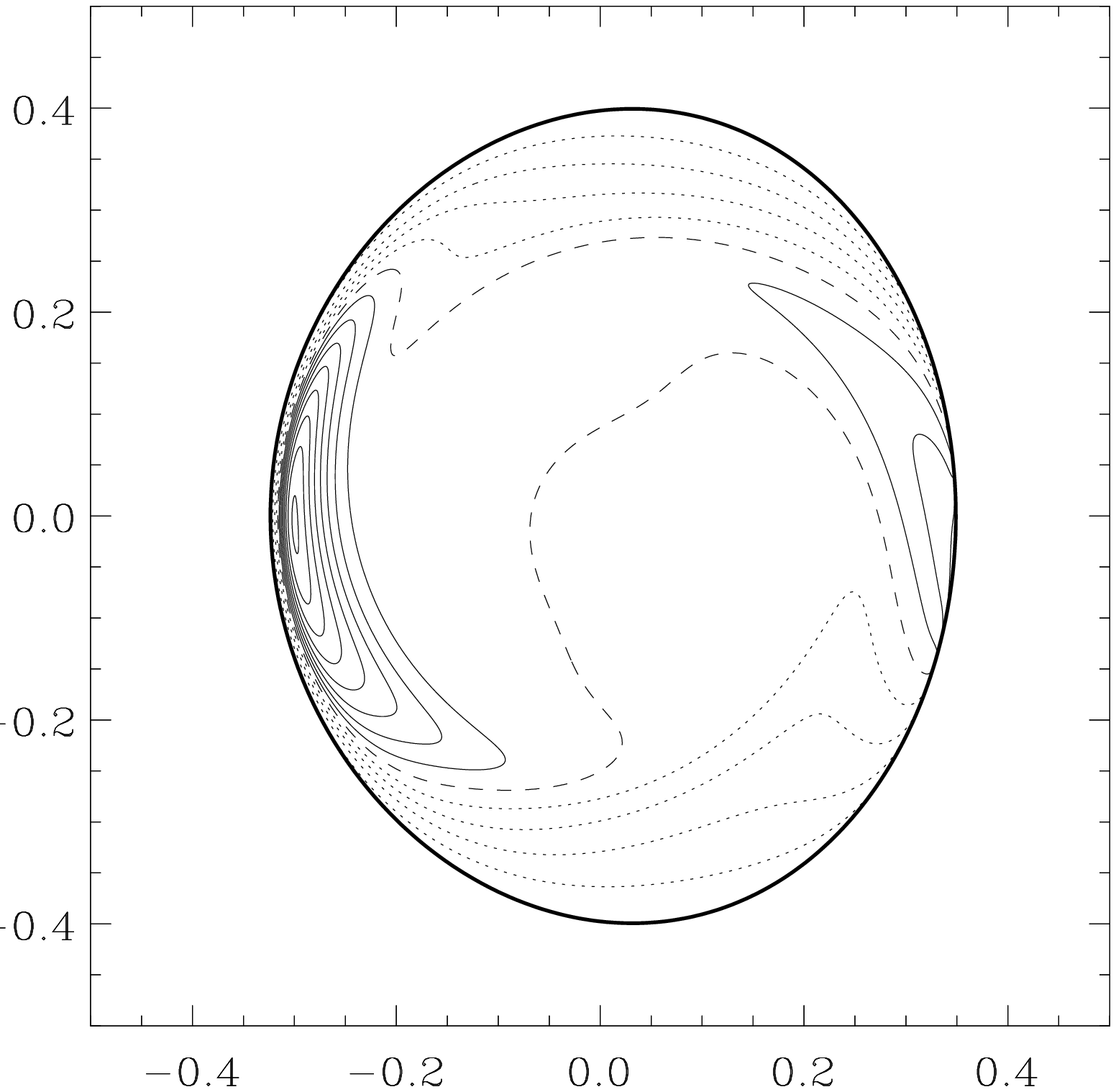}\qquad\epsfysize8cm\epsfbox{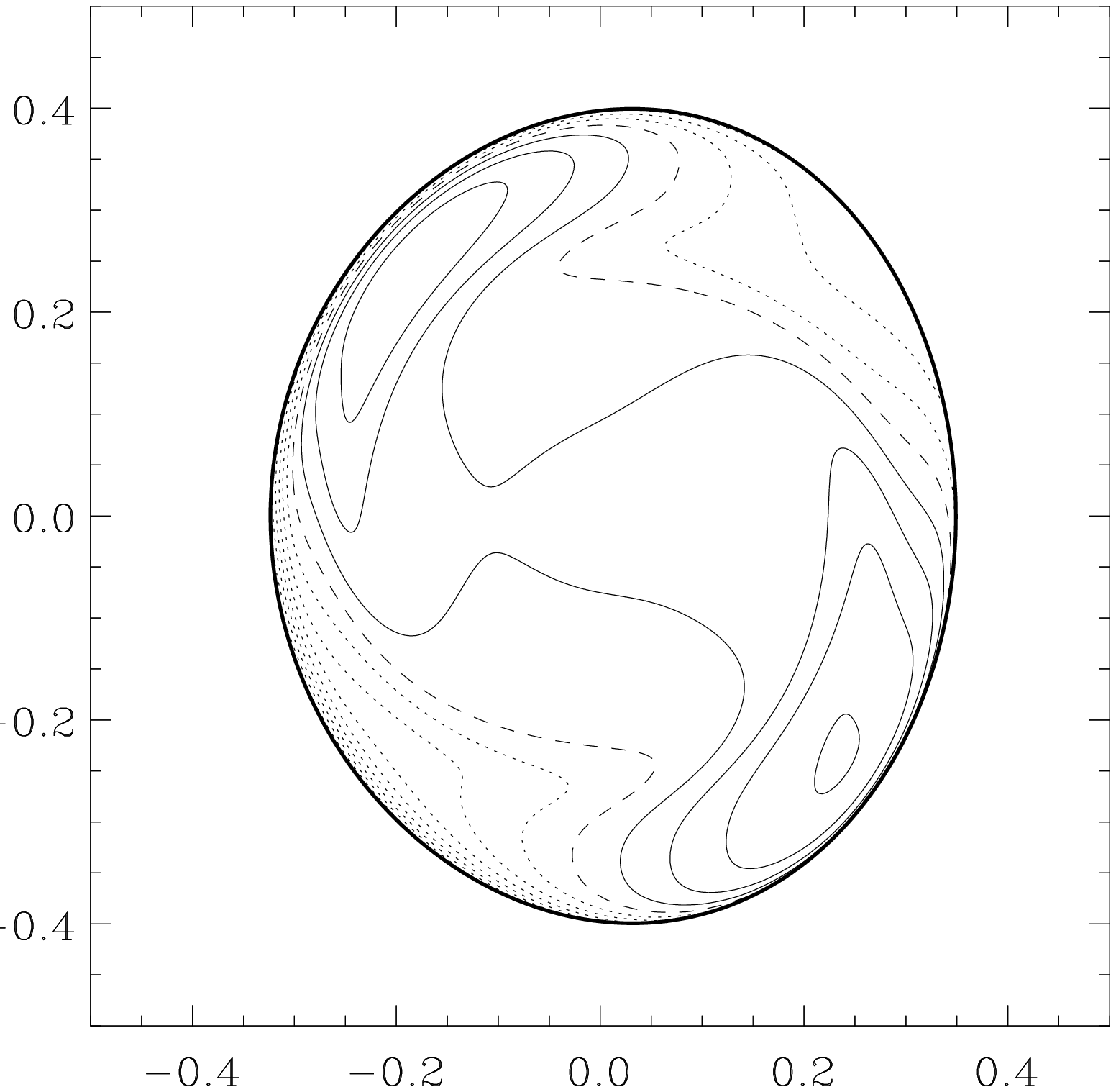}}
  \caption{Non-axisymmetric distortion of a disc with parameters
    $\alpha=0.1$, $\alpha_{\rm b}=0$, ${\rm We}=0.5$ and $\gamma=5/3$
    in a binary with $q=0.5$.  The axes are in units of the binary
    separation, with the companion located at $(1,0)$.  The thick line
    represents the outer edge of the disc.  Left: contours of the flux
    distortion $f_F$, with spacing $0.2$.  The dotted, dashed and
    solid contours represent values less than $1$, equal to $1$ and
    greater than $1$, respectively.  Right: contours of the height
    distortion $f_H$, with spacing $0.05$.}
\end{figure*}

These results are very typical.  Both the surface brightness and the
thickness display a dominant $m=2$ distortion in the outer part of the
disc.  The phases are different, however.  While the brightened
sectors of the outer disc are approximately aligned with the
companion, the thickened sectors are misaligned by approximately
$45\degr$.  When translated into velocity space, the brightened
sectors have the wrong phase to explain the tomograms of IP Peg and
other systems, while the thickened parts have the correct phase.

\begin{figure*}
  \centerline{\epsfysize7cm\epsfbox{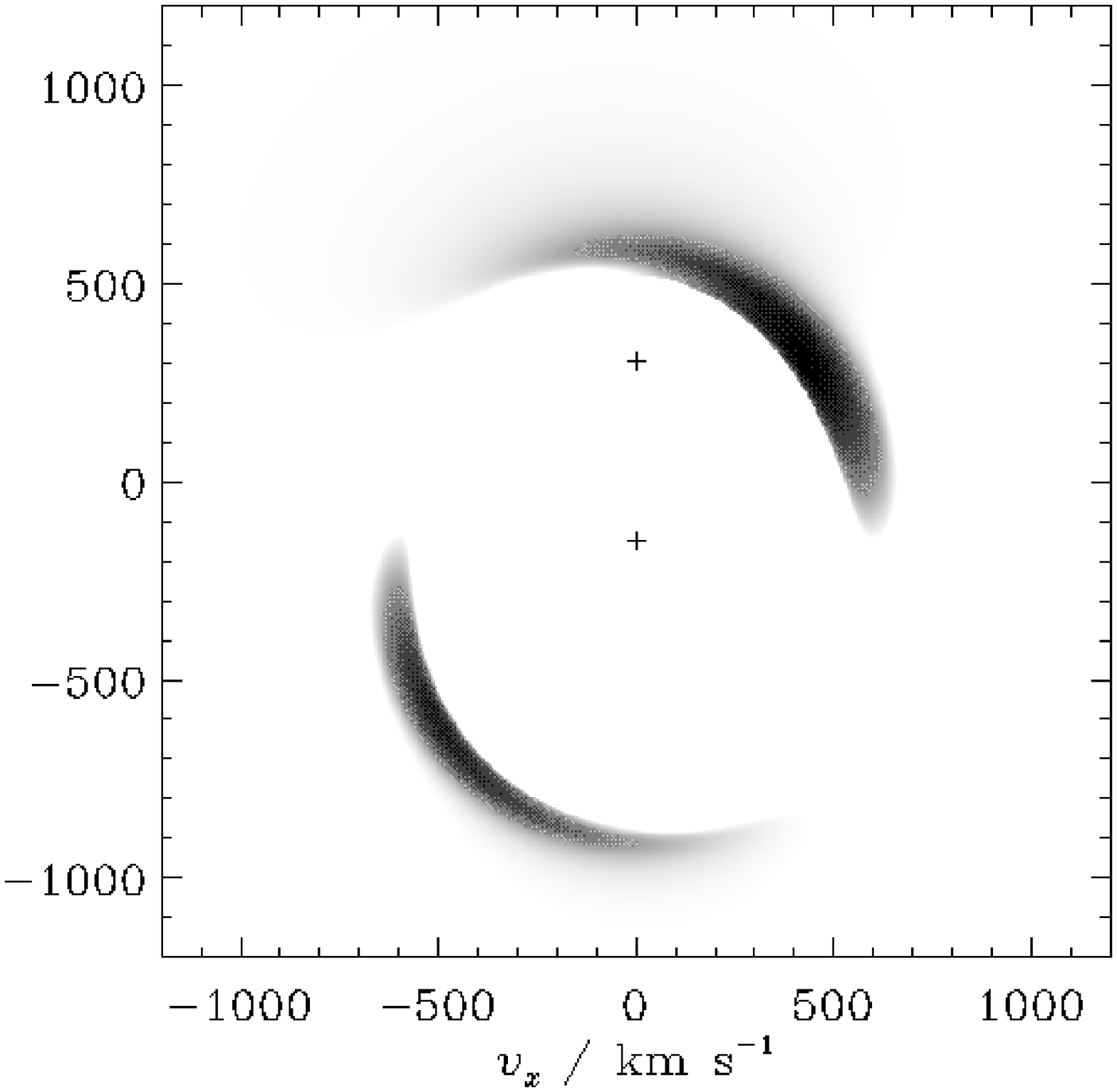}\qquad\epsfysize7cm\epsfbox{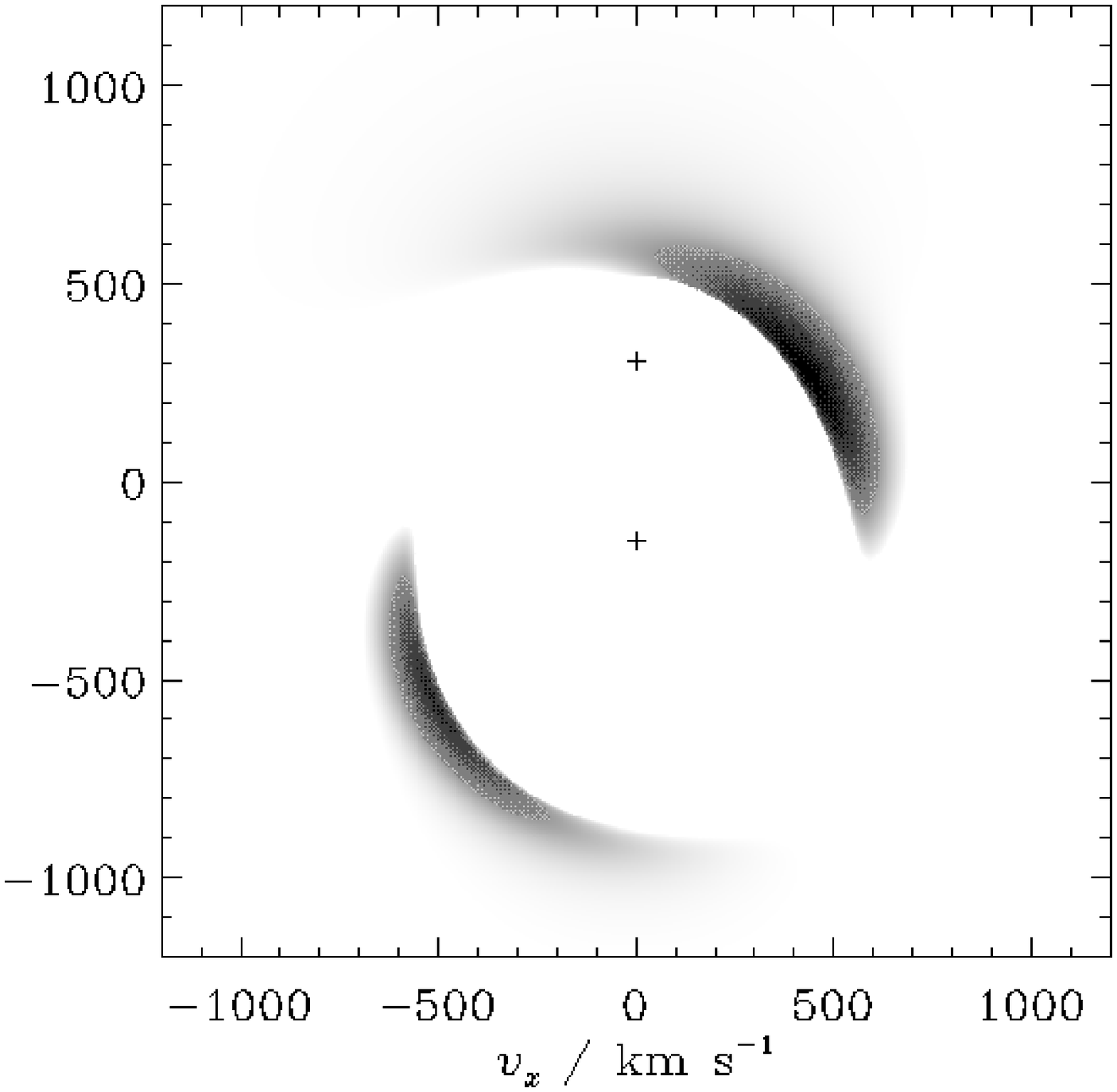}}
  \vskip0.3cm
  \centerline{\epsfysize7cm\epsfbox{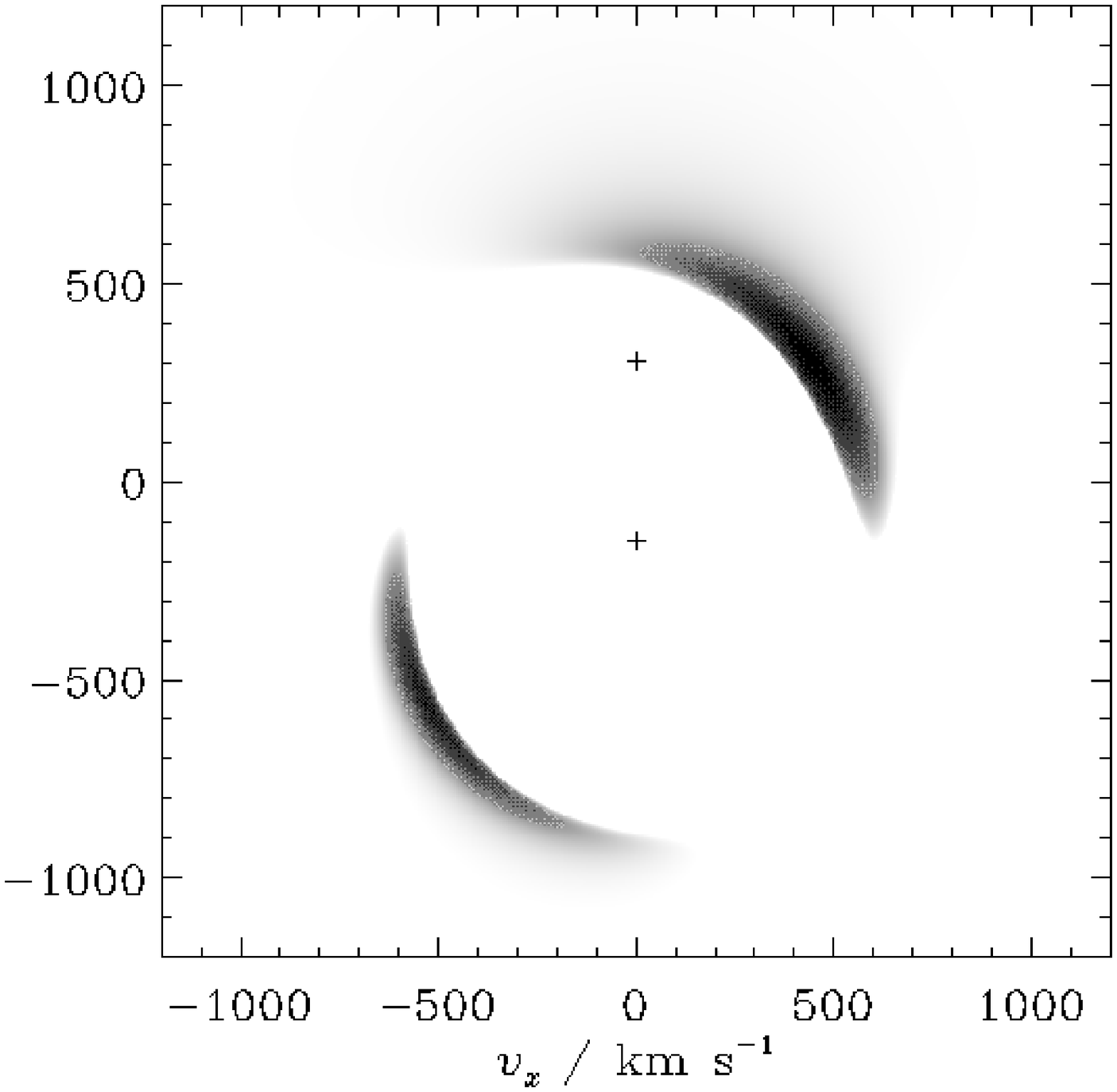}\qquad\epsfysize7cm\epsfbox{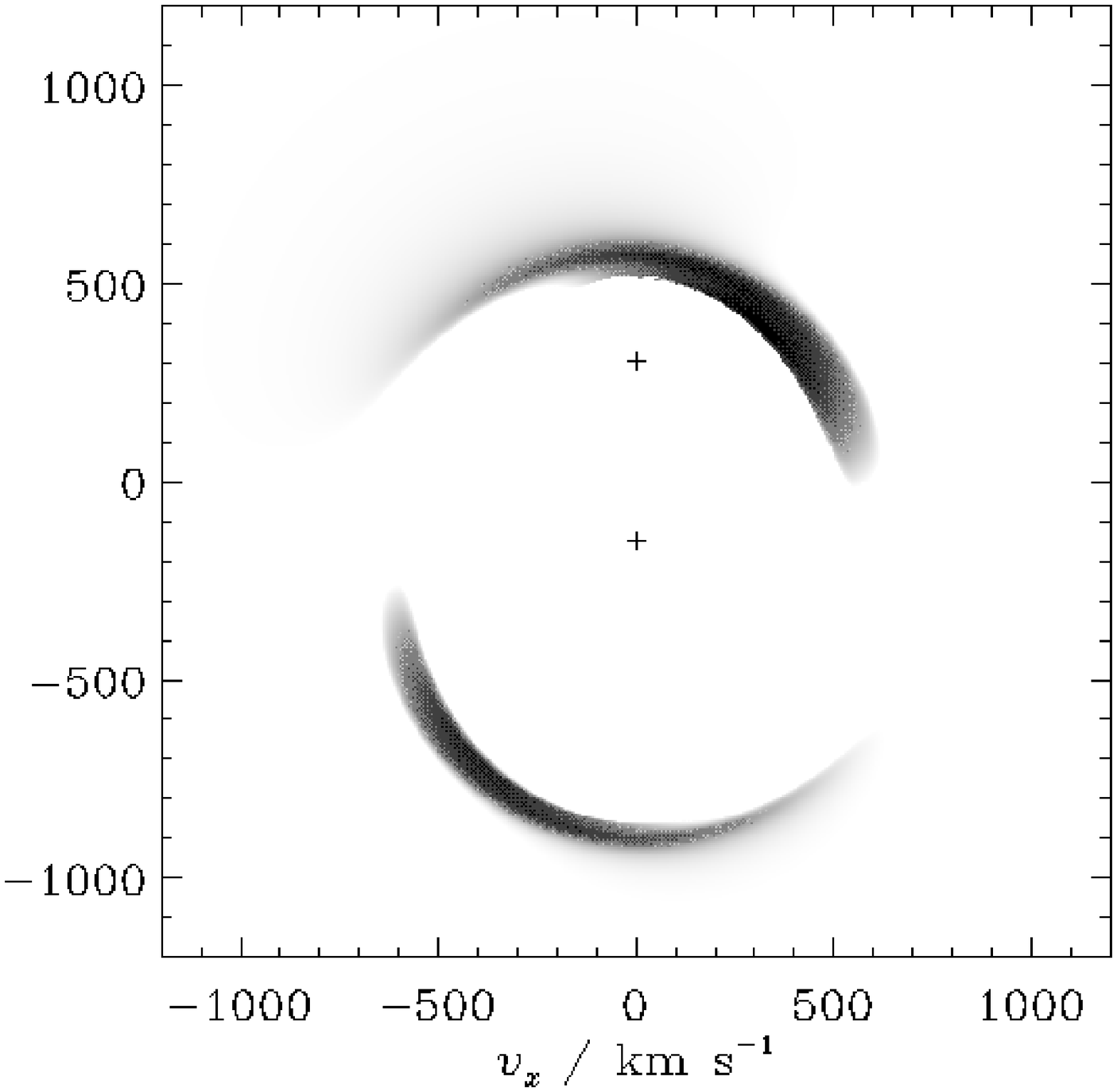}}
  \vskip0.3cm
  \centerline{\epsfysize7cm\epsfbox{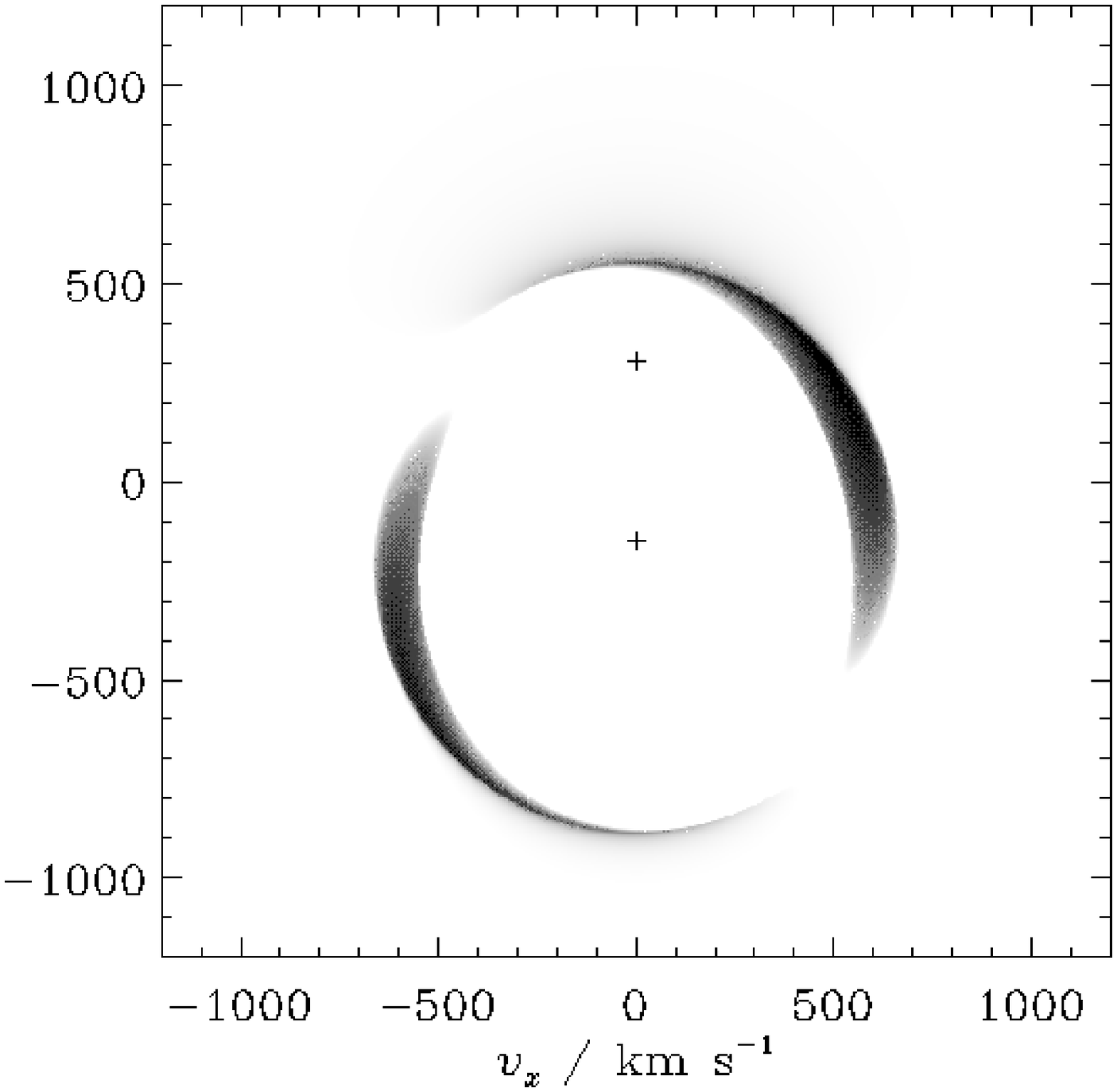}\qquad\epsfysize7cm\epsfbox{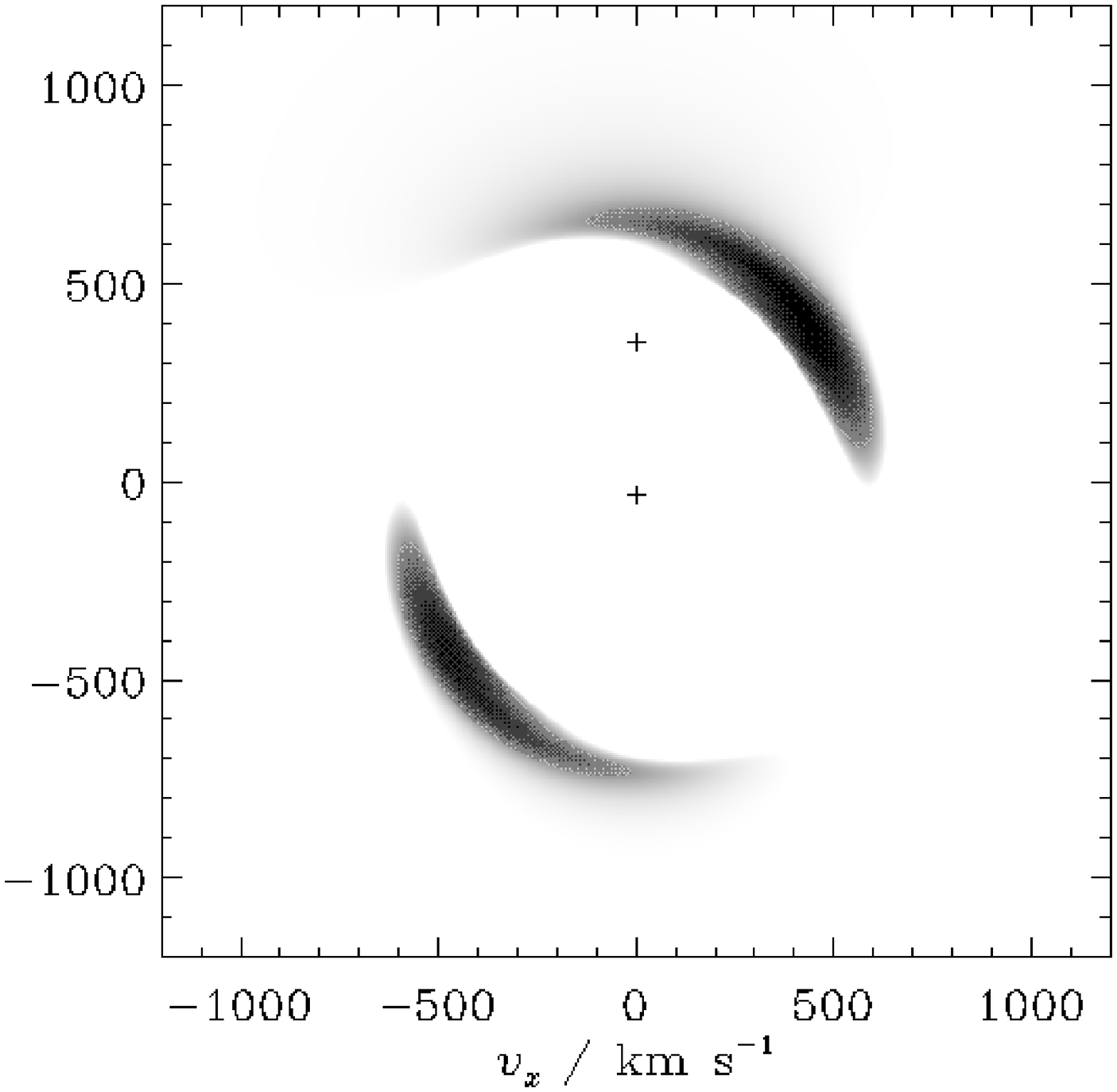}}
  \caption{Simple synthetic tomograms for a disc in a binary.  The
    velocity scale is based on the orbital parameters of IP Peg.  The
    plotted intensity is proportional to the elevation of the disc
    above its undistorted value.  Top left: standard parameters, as in
    Fig.~2.  Top right: standard except that $\alpha=0.3$.  Middle
    left: standard except that ${\rm We}=0.1$.  Middle right: standard
    except that ${\rm We}=1$.  Bottom left: standard except that
    $\gamma=1.2$.  Bottom right: standard except that $q=0.1$ (the
    velocity scale assumes the same value of $GM_1/a$ as in IP Peg).}
\end{figure*}

\subsection{Emission lines and synthetic Doppler tomograms}

This suggests that the observed patterns can be explained if the line
emission is correlated with the thickness of the disc rather than the
continuum brightness.  This is consistent with the notion that the
emission lines originate in parts of the disc that are elevated and
therefore irradiated from the centre of the system by the white dwarf,
boundary layer and/or inner disc (Robinson et al. 1993).

Simple synthetic Doppler tomograms based on this premise can then be
constructed by plotting an image representing the excess elevation of
the disc in velocity space.  The quantity plotted in Fig.~3 is
$f_H-f_{H0}$, where
\begin{equation}
  f_{H0}=\left(1+{{9}\over{4}}{\rm We}\,\alpha\right)^{13/28}
\end{equation}
is the constant value of $f_H$ in the absence of star 2.  The velocity
field used in constructing the tomograms is the orbital motion of
Paczy\'nski's orbits, translated into the inertial frame.

These synthetic tomograms should not be considered as a serious
attempt to reconstruct the observed data, such as that made by Steeghs
\& Stehle (1999).  This would require a better understanding of the
mechanism of line emission and a proper treatment of the instrumental
and data-analytical limitations of Doppler tomography.  Rather, Fig.~3
should be taken as an indication of the probable locations in velocity
space of line emission under the irradiation hypothesis.

The dependence of the two-armed patterns on the disc parameters is
rather subtle.  Increasing the values of $\alpha$, $\alpha_{\rm b}$ or
${\rm We}$ can cause phase shifts that rotate the patterns to some
extent.  Decreasing $\gamma$ moves the vertical resonance outwards in
real space (inwards in velocity space) and can also increase the
amplitude of the distortion by making the disc more compressible.

\subsection{The role of the vertical resonance}

\label{vr}

The significance of the vertical resonance in this problem can be
demonstrated by considering an approximate model that corresponds to a
linear, inviscid analysis of the full equations.  In the limit
$\alpha,\alpha_{\rm b}\to0$ the dimensionless turbulent stress
coefficients $t^{ab}$ tend to zero and one obtains the much simpler
system of ODEs
\begin{equation}
  g_\Omega\left(-\partial_\phi\ln f_1+
  {{13}\over{2}}\partial_\phi\ln f_2\right)=-14f_3-3g_1,
\end{equation}
\begin{equation}
  g_\Omega\left({{1}\over{\gamma-1}}\right)
  \partial_\phi\ln f_2=-\left({{\gamma+1}\over{\gamma-1}}\right)f_3-g_1,
\end{equation}
\begin{equation}
  g_\Omega\partial_\phi f_3=-f_3^2-g_5+f_2.
\end{equation}
When the tidal distortion is weak, one may write $f_2=1+\delta f_2$,
$g_\Omega=(1-g_\omega)+\delta g_\Omega$ and $g_5=1+\delta g_5$, where
$\delta f_2$, $f_3$, $\delta g_\Omega$, $g_1$ and $\delta g_5$ are
small quantities.  In the linear approximation, products of small
quantities may be neglected, and this leads to a linear, inhomogeneous
equation of second order for $\delta f_2$,
\begin{equation}
  \left[(1-g_\omega)^2\partial_\phi^2+(\gamma+1)\right]\delta f_2=X,
\end{equation}
where
\begin{equation}
  X=(\gamma+1)\delta g_5-(\gamma-1)(1-g_\omega)\partial_\phi\delta g_1.
\end{equation}
The forcing function may be resolved into Fourier components,
\begin{equation}
  X(\phi;\lambda)={\rm Re}\sum_{m=0}^{\infty}
  X_m(\lambda)\,{\rm e}^{{\rm i}m\phi},
\end{equation}
where the coefficients $X_m$ are real by virtue of the symmetry of
Paczy\'nski's orbits.  The solution is then
\begin{equation}
  \delta f_2={\rm Re}\sum_{m=0}^{\infty}
  \left[(\gamma+1)-m^2(1-g_\omega)^2\right]^{-1}X_m\,{\rm e}^{{\rm i}m\phi}.
\end{equation}
It is singular at the vertical resonances, where the quantity in
square brackets vanishes.  Now
\begin{equation}
  g_\omega=(1+q)^{1/2}\left({{\lambda}\over{a}}\right)^{3/2},
\end{equation}
and so the inner/outer resonance for mode $m>0$ occurs at
\begin{equation}
  {{\lambda}\over{a}}=(1+q)^{-1/3}\left[1\mp{{1}\over{m}}
  (\gamma+1)^{1/2}\right]^{2/3},
\end{equation}
as determined by Lubow (1981).  In particular, for $q=0.5$ and
$\gamma=5/3$, the $m=2$ inner vertical resonance occurs at
$\lambda/a\approx0.282$.

The phase of the response to $m=2$ forcing changes by $\pi$ as one
passes through the radius of the resonance.  In the linear, inviscid
limit, the response is formally singular at the resonance.  In
reality, the response is limited by the emission of a wave (Lubow
1981), or by non-linearity or dissipation (Ogilvie, in preparation).
The singularity is then avoided and a smooth variation of amplitude
and phase occurs.

The distortion of the thickness of the disc is sensitive mainly to the
quantity $f_2$, according to equation (\ref{fh}).  Therefore the
vertical distortion is amplified in the vicinity of the vertical
resonance and undergoes a phase shift as one passes through the
resonance.  This effect can be seen in Fig.~2 (right panel).  For the
$m=2$ component of the solution, the phase shift of $\pi$ corresponds
to a rotation of the peak through $90\degr$.  This lends a certain
`spirality' to the pattern in the vicinity of the resonance.  However,
the solution does not represent a wave or shock, and the shape of the
pattern is independent of the thickness of the disc.

\section{Summary and discussion}

In this paper I have constructed a detailed semi-analytical model of
the non-linear tidal distortion of a thin, three-dimensional accretion
disc by a binary companion on a circular orbit.  The analysis allows
for vertical motion and radiative energy transport, and introduces a
simple model for the turbulent magnetic stress.  The fluid dynamical
equations are solved by means of a consistent asymptotic expansion,
using the fact that the disc is thin.  The solution is formally exact
in the limit of a thin disc.
  
The tidal distortion affects quantities such as the surface brightness
of the disc and its thickness.  As expected, the distortion is
greatest in the outer part of the disc and is predominantly two-armed.
An important feature of the solution is the $m=2$ inner vertical
resonance, first analysed by Lubow (1981), which typically lies within
the tidal truncation radius in binary stars.  This intrinsically
three-dimensional effect has a significant influence on the amplitude
and phase of the tidal distortion.

I have shown that the two-armed features observed in Doppler tomograms
of the discs of dwarf novae in outburst find a natural explanation if
the emission is associated with the tidally thickened sectors of the
outer disc that are elevated and may therefore be irradiated by the
white dwarf, boundary layer and/or inner disc during the outburst.
There is a slight spirality to the structure, arising from the phase
shift that occurs on passing through the vertical resonance.  There is
some dependence of the pattern on the mass ratio and the dimensionless
parameters $(\alpha,\alpha_{\rm b},{\rm We},\gamma)$ of the disc
(defined in Section~\ref{basic}).  In particular, the adiabatic
exponent $\gamma$ determines the location of the vertical resonance.
Future observations of higher quality might help to constrain the
values of these parameters.  In principle, the distortion of the
continuum surface brightness of the disc, which displays a different
pattern (e.g. Fig.~2), might also be seen in eclipse-mapping
observations.

This possible interpretation of the observed features is fundamentally
different from one based on spiral waves or shocks.  There are no
waves or shocks in the present analysis.  The pattern of emission
merely reflects the tidal distortion of the disc and is independent of
the angular semi-thickness $H/r$, in the limit that this is small.
The observability of the pattern will nevertheless vary during the
outburst cycle depending on the spreading of mass to the outer parts
of the disc and, in the irradiation hypothesis, on the luminosity of
the central regions.

In a recent paper, Smak (2001) has presented arguments against the
interpretation of the observed features as spiral shocks.  He argues
in favour of the irradiation hypothesis and suggests that sectors of
the outer disc may be elevated through the horizontal convergence of
the three-body orbits.  The present paper supports the arguments of
Smak (2001) to a certain extent, but within the context of a
three-dimensional fluid dynamical model.  The principal factors
leading to the thickening of the disc can be seen in Section~\ref{vr}.
Vertical expansions and contractions are driven by the variation of
the vertical gravitational acceleration around the tidally distorted
orbits and, to some extent, by the horizontal convergence of the flow
in the binary plane.  The response of the disc to this forcing is
regulated by the inner vertical resonance.

The analysis in this paper could be extended and improved in a number
of ways.  Mass input to the disc, its ultimate tidal truncation, and
the spreading of mass during an outburst, should properly be discussed
within the context of the evolutionary equation for the surface
density, alluded to in Section~\ref{slow} but not derived here.  The
solution of this equation would determine the relative weighting to be
assigned to the different orbits.  The viscoelastic model for the
turbulent stress, although an improvement on purely viscous
treatments, could also be refined.  Finally, a better understanding of
the mechanism of line emission would enable a closer connection to be
made between the computed models and the observational data.

\section*{Acknowledgments}

I thank Steve Lubow for many helpful discussions at an early stage in
this investigation, Jim Pringle for constructive comments on the
manuscript, and Alex Schwarzenberg-Czerny for bringing the work of
Smak (2001) to my attention.  I acknowledge the support of the Royal
Society through a University Research Fellowship.

\label{lastpage}

\end{document}